\ifdefined\DocumentMetadata
  \DocumentMetadata{}
\fi
\documentclass[acmsmall,screen,nonacm]{acmart}

\usepackage[subrefformat=parens]{subcaption}
\usepackage{braket}
\usepackage[ruled,vlined,linesnumbered]{algorithm2e}
\usepackage{multicol,multirow}
\makeatletter
\DeclareRobustCommand\origttfamily
{\not@math@alphabet\ttfamily\mathtt
\fontfamily\ttdefault\selectfont\hyphenchar\font=-1}
\DeclareTextFontCommand{\texttorig}{\origttfamily}
\DeclareRobustCommand\ttfamily
{\not@math@alphabet\ttfamily\mathtt
\fontfamily\ttdefault\selectfont\hyphenchar\font=`\-}
\makeatother

\SetKwComment{Comment}{$\triangleright$\ }{}
\DontPrintSemicolon
\newcommand{\algoname}[1]{\textnormal{\textsc{#1}}}

\AtBeginDocument{
  \providecommand\BibTeX{{
    \normalfont B\kern-0.5em{\scshape i\kern-0.25em b}\kern-0.8em\TeX}}}

\setcopyright{cc}
\setcctype{by}
\acmPrice{}
\copyrightyear{2025}
\acmYear{2025}
\acmDOI{10.1145/3748261}

\acmJournal{TQC}
\acmVolume{6}
\acmNumber{4}
\acmArticle{27}
\acmArticleSeq{0}
\acmMonth{8}

\begin{document}

\title{Lazy Qubit Reordering for Accelerating Parallel State-Vector-based Quantum Circuit Simulation}
\thanks{\textcopyright~2025 Copyright held by the owner/author(s).
\emph{ACM Transactions on Quantum Computing} 6(4), Article 27, 1--33 (2025), \url{https://doi.org/10.1145/3748261}.}

\author{Yusuke Teranishi}
\affiliation{
 \institution{Graduate School of Information Science and Technology, The University of Osaka}
 \streetaddress{1-5 Yamadaoka}
 \city{Suita}
 \state{Osaka}
 \postcode{565-0871}
 \country{Japan}}
\email{u164648h@alumni.osaka-u.ac.jp}

\author{Shoma Hiraoka}
\affiliation{
 \institution{Graduate School of Information Science and Technology, The University of Osaka}
 \streetaddress{1-5 Yamadaoka}
 \city{Suita}
 \state{Osaka}
 \postcode{560-0871}
 \country{Japan}}
\email{u564252i@ecs.osaka-u.ac.jp}

\author{Wataru Mizukami}
\affiliation{
 \institution{Center for Quantum Information and Quantum Biology, The University of Osaka}
 \streetaddress{1-2 Machikaneyama}
 \city{Toyonaka}
 \state{Osaka}
 \postcode{560-0043}
 \country{Japan}}
\email{mizukami.wataru.qiqb@osaka-u.ac.jp}

\author{Masao Okita}
\affiliation{
 \institution{Graduate School of Information Science and Technology, The University of Osaka}
 \streetaddress{1-5 Yamadaoka}
 \city{Suita}
 \state{Osaka}
 \postcode{565-0871}
 \country{Japan}}
\email{okita@ist.osaka-u.ac.jp}

\author{Fumihiko Ino}
\affiliation{
 \institution{Graduate School of Information Science and Technology, The University of Osaka}
 \streetaddress{1-5 Yamadaoka}
 \city{Suita}
 \state{Osaka}
 \postcode{565-0871}
 \country{Japan}}
\email{ino@ist.osaka-u.ac.jp}

\renewcommand{\shortauthors}{Teranishi et al.}

\begin{abstract}
This article proposes two quantum operation scheduling methods for accelerating parallel state-vector-based quantum circuit simulation using multiple graphics processing units (GPUs).
The proposed methods reduce all-to-all communication caused by qubit reordering, which can dominate the overhead of parallel simulation.
Our out-of-order approach eliminates redundant reorderings by introducing intentional delays in reordering communications such that multiple reorderings can be aggregated into a single reordering.
The delays are carefully introduced based on the principles of time-space tiling, or a cache optimization technique for classical computers, which we use to arrange the execution order of quantum operations.
Moreover, we develop these methods tailored for two primary procedures in variational quantum eigensolver simulation: quantum state update (QSU) and expectation value computation (EVC).
Our QSU simulation takes an advantage of the hierarchical interconnection of GPU systems to avoid slow inter-node communication.
On the other hand, our EVC simulation reduces the number of reorderings by diagonalization of Pauli strings.
Experimental validation on 32-GPU executions demonstrates acceleration in QSU and EVC --- up to 54$\times$ and 1,657$\times$, respectively --- compared to an inorder-based method.
We believe that our out-of-order approach is useful for accelerating large-scale quantum circuit simulations, including QSU and/or EVC that operate qubits in a regular manner.
\end{abstract}



\received{29 September 2024}
\received[revised]{3 May 2025}
\received[accepted]{27 June 2025}

\maketitle

\section{Introduction}
Quantum circuit simulations on classical computers are crucial in validating quantum algorithms intended for quantum computers.
While recent years have seen the appearance of quantum circuit simulators~\cite{qulacs,qiskit,cuquantum} compatible with commodity computers, the primary limitation lies in the main memory capacity, restricting the number of qubits in state-vector-based quantum circuit simulation (SVQCS)~\cite{qulacs,qiskit,cuquantum}; 16GB of memory is needed to store a state vector for 30 qubits in double precision, for example.
With the increasing number of qubits in real quantum computers, there is a growing need to conduct large-scale quantum circuit simulations on classical computers.
Moreover, to explore quantum supremacy, it is essential to develop highly efficient simulation methods leveraging noiseless classical computers as a benchmark against real quantum computers.

Parallel SVQCS (p-SVQCS)~\cite{p-svqcs} is a promising approach for scaling up simulations to accommodate a larger number of qubits ($>30$).
Central to p-SVQCS is the distribution of the state vector across multiple computers, enabling data parallelism in quantum gates, \textit{i.e.}, quantum operations.
With an appropriate distribution of the state vector, individual computers can reduce computation time by applying quantum operations to their respective partial state vectors in a data-parallel manner.
However, the appropriate distribution depends on the quantum operation, so that qubit reordering \cite{p-svqcs,cuquantum} is necessary during p-SVQCS to realize better distribution that fits to subsequent operations.

Efficient large-scale p-SVQCS encounters challenges in achieving high parallel efficiency, primarily owing to the frequent occurrence of qubit reorderings, which significantly increase the communication time due to all-to-all communication.
For instance, our experimantal results show that the cumulative communication time for qubit reorderings comprises 60\% of the total execution time on 8 graphics processing units (GPUs) that simulate a 34-qubit GateFabric circuit~\cite{gatefabric}.
Addressing this issue is paramount for enhancing the efficiency of p-SVQCS~\cite{mpi-qulacs,qiskit-aer} with the best tradeoff point between computation and communication.
Existing solutions such as mpiQulacs~\cite{mpi-qulacs} focused on reducing communication time by aggregating successive qubit reorderings, while Qiskit Aer~\cite{qiskit-aer} introduced heuristic scheduling methods to reduce the number of qubit reorderings during simulations.
These efforts underscore the importance of minimizing qubit reordering occurrences in p-SVQCS, but the existing solutions deploy an inorder approach, which limits the impact of qubit reordering.

We propose two novel scheduling methods tailored for p-SVQCS to reduce the frequency of qubit reorderings, on the subject of the variational quantum eigensolver (VQE)~\cite{vqe}, a representative quantum-classical hybrid algorithm.
Our out-of-order methods change the execution order of quantum operations, particularly within quantum state update (QSU) and expectation value computation (EVC) --- key components of VQE.
Central to our approach is intentionally lazy qubit reorderings, which increase computational granularity per communication, inspired by the principles of time-space tiling ~\cite{time_space_tiling}, or a cache optimization technique for stencil computation~\cite{stencil_computation}.
This deliberate delay, which facilitates the aggregation of multiple qubit reorderings into a single qubit reordering, is useful to eliminate redundant qubit reorderings.
We also present an extended scheduling method tailored for QSU, focusing on the hierarchical interconnection of GPU cluster systems.
This method avoids slow inter-node communication to reduce overall communication time, albeit at the cost of a slightly increased number of qubit reorderings.
Furthermore, we present an optimization method for EVC, which reduces the occurence of qubit reorderings.
We implemented these methods for p-SVQCS on multiple GPUs using NVIDIA cuQuantum~\cite{cuquantum}.
Our implementation is freely available at https://github.com/qiqb-osaka/chemqulacs-gpu/.

This study offers several significant contributions.
\begin{enumerate}
	\item Demonstrating the usefulness of time-space tiling in enhancing p-SVQCS. Time-space tiling is useful for realizing our out-of-order approach, which realizes more aggregations of qubit reorderings than inorder approaches.
	\item Introducing an efficient qubit reordering-based QSU approach tailored for hierarchical interconnection. Our approach assumes that intra-node communication is faster than inter-node communication. Based on this assumption, the approach accepts the increased number of qubit reorderings to reduce overall communication time. Such a solution cannot be realized by the existing methods because more qubit reorderings result in longer communication time in flat interconnection.
	\item Proposing a qubit reordering-based parallel EVC approach coupled with optimal scheduling of quantum operations, namely Pauli strings. In contrast to existing inorder methods, the proposed method is optimized by out-of-order reordering of qubits, tiling of Pauli strings, and diagonalization of Pauli strings, which mitigates dependencies between computation and communication to facilitate scheduling of quantum operations.
\end{enumerate}
We believe that the proposed methods extend its utility beyond the VQE, proving beneficial for various quantum algorithms involving QSU and/or EVC that operate qubits in a regular manner.

\section{Related Work}
Various studies have been conducted to accelerate p-SVQCS, focusing on the communication process~\cite{p-svqcs,qiskit-aer,qiskit-aer-noise,cuquantum-appliance,comm_optimize,hq-sim,mpi-qulacs,quest,intel-qs,qhipster,hyquas,sv-sim,shorgpu,tyson}.
These methods include gate scheduling for reducing qubit reordering occurrences~\cite{p-svqcs,qiskit-aer,qiskit-aer-noise,cuquantum-appliance,comm_optimize}, computation-communication overlap for hiding communication~\cite{quest,intel-qs,qhipster,hyquas,sv-sim}, and specialized communication for specific quantum operations~\cite{shorgpu,tyson}.
Most of them adopt a method that exchanges distributed qubits, \textit{i.e.}, local and global qubits, by performing all-to-all communication for qubit reordering.
Regarding the gate scheduling approach, existing proposals rely on heuristic methods due to the combinatorial explosion inherent in the scheduling problem.
In other words, it is unclear how to find the optimal schedule, which minimizes simulation time with the best execution order of gates and the best assignment of local and global qubits.
This problem resembles quantum circuit mapping~\cite{quantum_circuit_mapping}, which compiles a high-level quantum circuit model to satisfy the physical constraints of low-level real quantum hardware.
However, finding the optimal quantum circuit mapping is known to be an NP-complete problem~\cite{quantum_circuit_mapping,qcp_np_1,qcp_np_2}.

Several gate scheduling methods~\cite{p-svqcs,qiskit-aer,qiskit-aer-noise,cuquantum-appliance,comm_optimize} was proposed for QSU.
H\"{a}ner and Steiger~\cite{p-svqcs} were pioneers who realized p-SVQCS for a 45-qubit quantum supremacy circuit.
Their general method reorders the gates and exchanges the global qubits for the lowest local qubits selected by an efficient search algorithm.
Jiao et al.~\cite{comm_optimize} proposed a gate-aware on-demand communication method, which exploits the data locality between quantum gates and avoids unnecessary inter-node communications.
Such simple approaches are easy to implement but may fail to reduce qubit reorderings; reordering inappropriate qubits cause an increase in qubit reorderings, which we avoid by selecting qubits locally optimal.
Doi et al.~\cite{qiskit-aer,qiskit-aer-noise} developed Qiskit Aer, which heuristically finds a suitable schedule that includes less qubit reorderings.
Their scheduling method requires a time complexity of $\mathcal{O}(m^2)$, where $m$ is the number of gates in the simulated circuit.
Therefore, the approach can suffer from long scheduling time for large $m$; the UCCSD circuit~\cite{uccsd,uccsd_resources} has $m > 10^5$, for example.
NVIDIA developed cuQuantum Appliance~\cite{cuquantum-appliance}, extending the backend capabilities of Qiskit Aer~\cite{qiskit-aer} to integrate with the cuQuantum library~\cite{cuquantum}.
This solution demonstrated simulations of 41 qubits on 64 nodes (512 GPUs) in the AIST ABCI system~\cite{custatevec_abci, custatevec_benchmark}.
However, the scheduling methodology remains inaccessible because the source code is encapsulated within a Docker container.
Our method also deploys a heuristic method but differs from the previous methods in terms of integrating the principles of time-space tiling, which realizes efficient qubit reorderings with reduced communication cost, albeit at the cost of a slight time complexity.
The scheduling overhead is negligible compared to the subsequent QSU simulation because the former takes less time than the latter: QSU simulation for $n$ qubits has an exponential time complexity of $n$ whereas the proposed scheduling algorithm for QSU has a time complexity of $\mathcal{O}(\overline{d}m)$, where $\overline{d}$ is the average number of target qubits.

An important acceleration approach is to reduce the communication time per qubit reordering.
Zhang et al.~\cite{hq-sim} reduced communication overhead by lossy data compression, which can reduce the amount of data transfer.
Tabuchi et al.~\cite{mpi-qulacs} developed mpiQulacs, a parallel version of Qulacs~\cite{qulacs} optimized for the Fujitsu A64FX CPU cluster system.
They reduced the amount of inter-node and intra-node communication by aggregating successive reorderings of a single qubit into a single reordering of multiple qubits.
Suppose that $k$ reorderings of 1 qubit can be replaced with a single reordering of $k$ qubits.
This replacement reduces the amount of communication by $(k/2-1+2^{-k}) 2^n$ because the amount of data transfer for a single reordering of $k$ qubits is given by $(1-2^{-k}) 2^n$.
In contrast to their inorder approach, our out-of-order method changes the execution order of quantum operations such that more multiple communications can be merged into a single communication.

Another acceleration approach is to realize computation-communication overlap~\cite{quest,intel-qs,qhipster,hyquas,sv-sim}.
The Intel-QS~\cite{intel-qs} (also known as \textit{q}H\textit{i}PSTER~\cite{qhipster}) and QuEST~\cite{quest} simulators perform a qubit reordering while applying quantum gates to a state vector.
Zhang et al.~\cite{hyquas} presented a GPU-centric pipelining approach, which simultaneously perform all-to-all communication and a gate application to hide the communication latency.
Li et al.~\cite{sv-sim} proposed a partitioned global address space (PGAS) based state-vector simulator, which realizes fine-grained computation-communication overlap by leveraging the SHMEM communication model~\cite{shmem}.
Our methods can be integrated with these studies.

Some researchers~\cite{shorgpu,tyson} reduced communication frequency for specific quantum operations.
Willsch et al.~\cite{shorgpu} proposed a communication process specializing in quantum operations that appear in Shor's algorithm~\cite{nielsen}: the rotation and the controlled modular multiplication.
Jones et al.~\cite{tyson} precisely assessing qubit reordering frequency by analyzing memory access and communication patterns for typical quantum operations.
In contrast to these gate-specific approaches, our general approach aims to reduce the qubit reordering frequency for two-qubit gates.
Our general approach can be integrated with the gate-specific approaches to further reduce qubit reordering occurrences.

As for GPU-based parallelization, simulators such as CUDA Quantum~\cite{cudaq,cudaq_blog} and Pennylane Lightning~\cite{pennylane_lightning} (plugins of Pennylane~\cite{pennylane}) parallelize EVC by dividing the computation in terms of Hamiltonian terms and assigning the divided parts to GPUs.
However, these simulators assume that GPUs hold the entire set of qubits.
Therefore, the maximum number of simulated qubits is strictly limited by the memory capacity of a single GPU.
Conversely, our qubit reordering-based parallel algorithm distributes the state vector to computing nodes, which allows increasing the number of qubits for large qubit systems.
We empirically found that our optimal tiling method can estimate the expectation values of $O(n^4)$ terms in the molecular Hamiltonian in just three qubit reorderings.

To the best of our knowledge, this is the first efficient parallel implementation utilizing qubit reordering for EVC.
Moreover, our extended method maximizes the performance on the hierarchical interconnection of GPU cluster systems by utilizing fast intra-node communication rather than slow inter-node communication.
Such a hierarchical interconnection can be found in non-uniform memory access (NUMA) systems, where CPU cores can access local memory and non-local memory in the same node.
As for the impact to quantum computing, the proposed methods can localize the communication on large-scale quantum systems that distribute qubits to multiple nodes. We believe that the time-space tiling is also useful for distributed quantum systems, where quantum processing units can generate intra-node and inter-node communication.

\section{Preliminaries}

We describe general quantum computation with a 3-tuple $(Q,C,T)$, named the QCT form,
where $Q=\{0,1,\ldots,n-1\}$ represents the set of $n$ qubits available in the quantum system, $C=\{C_0,C_1,\ldots,C_{m-1}\}$ represents a partially ordered set of $m$ quantum operators, and $T: C\rightarrow 2^Q$ is a target map between $C$ and the target qubit of each operator.
The partial order for $C$ is determined by data dependency $\succ$, which is a homogeneous relation on $C$.
Notably, $C_i\succ C_j$, where $0 \leq i,j <m$, represents that $C_j$ directly depends on $C_i$.
An operator $C_i \in C$ affects a subset of $Q$; we denote this subset as the operator's target.

In this study, we adopt a state-vector method~\cite{statevector_sim,nielsen} for simulating quantum circuits.
A state vector $\ket{\psi}$ comprises $2^n$ complex elements:
\begin{align}
    \ket{\psi}=
    \begin{bmatrix}
        c_{0\ldots00} \\ c_{0\ldots01} \\ \vdots \\ c_{1\ldots11}
    \end{bmatrix},
    \label{eq:sv}
\end{align}
where $c$ is the probability amplitudes whose subscript, or an $n$-digit binary index, represents the alignment of the qubits in the simulator.
Without loss of generality, we assume that the simulator initially aligns the qubits in ascending order, with the least and most significant bits in the index corresponding to qubits $0$ and $n-1$, respectively.
In state-vector methods, each quantum gate is represented by a $2^n \times 2^n$ sparse matrix, and thus, SVQCS involves matrix-vector multiplication.

To establish the execution order of quantum operations in a simulation, we introduce the schedule $g$ as a sequence of operations, represented as a bijective map $g:\{0,1,\ldots,m-1\}\rightarrow C$; $g(x)=C_j$ indicates that operator $C_j$ must be processed as the $x$-th operation in the simulation.
In QSU, the schedule $g$ has to satisfy a constraint as follows:
\begin{gather}
    \forall C_i \in C, \forall C_j \in C\,[C_i \overset{+}{\succ} C_j \Rightarrow g^{-1}(C_i) < g^{-1}(C_j)], \label{eq:gate_order_constraint}
\end{gather}
where $\overset{+}{\succ}$ represents a transitive closure of $\succ$; $C_i \overset{+}{\succ} C_j$ requires that $C_i$ has to be computed before $C_j$. Hereafter, let C1 denote the constraint mentioned above.

\subsection{Variational Quantum Eigensolver}

A VQE~\cite{vqe} approximates the smallest eigenvalue of an eigenvalue equation, such as the Schrödinger equation.
The following equation represents a VQE designed to find the smallest eigenvalue $E$ of the Hamiltonian $H$ using ansatz $U(\vec{\theta})$:
\begin{align}
    E=\min_{\vec{\theta}} \bra{\psi_0}U^{\dagger}(\vec{\theta}) \,H\, U(\vec{\theta})\ket{\psi_0}, \label{eq:vqe}
\end{align}
where $\ket{\psi_0}$ is the initial state vector and $\vec{\theta}$ is the vector that contains the variational parameters.
A conventional VQE approach iteratively computes the QSU and EVC with an optimizer that updates parameter $\vec{\theta}$.
The quantum computations involved in QSU and EVC can be described by Eqs.~\eqref{eq:vqe_qsu} and \eqref{eq:vqe_evc}, respectively:
\begin{align}
    \ket{\psi(\vec{\theta})}=U(\vec{\theta})\ket{\psi_0}, \label{eq:vqe_qsu}\\
    E(\vec{\theta})=\bra{\psi(\vec{\theta})}H\ket{\psi(\vec{\theta})},\label{eq:vqe_evc}
\end{align}
where $E(\vec{\theta})$ is a variational expectation value of $H$.

In this study, we conduct both QSU and EVC through quantum simulation on classical computers.
The summation of simulation times for QSU and EVC, which depend on the size of parameters and that of the Hamiltonian, respectively, dictates the total execution time for the VQE.
Other components, such as parameter selection time, are typically negligible.

\subsubsection{Quantum state update}

\begin{figure}[t]
    \centering
    \includegraphics[height=30mm]{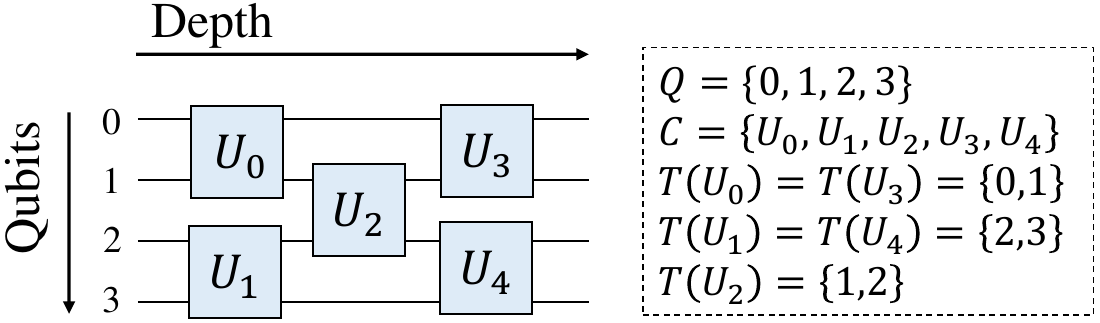}
    \caption{Example of a quantum circuit for a QSU and its QCT form.
    The dependencies of the gates are as follows: $U_0 \succ U_2$, $U_1 \succ U_2$, $U_2 \succ U_3$, and $U_2 \succ U_4$ result in $U_0 \overset{+}{\succ} U_3$, $U_1 \overset{+}{\succ} U_3$, $U_2 \overset{+}{\succ} U_3$, $U_0 \overset{+}{\succ} U_4$, $U_1 \overset{+}{\succ} U_4$, and $U_2 \overset{+}{\succ} U_4$.
    }
    \Description[Quantum circuit and its QCT representation.]{This is a quantum circuit diagram with four qubits and five two-qubit gates (U0 to U4). A box on the right provides the circuit's representation in QCT (Qubits, Operators, Target map) form. Q is the set of qubits {0,1,2,3}, C is the set of operators {U0 to U4}, and T specifies the target qubits for each operator.}
    \label{fig:qsu_qct}
\end{figure}

QSU obtains the final state $\ket{\psi}$ by applying the provided quantum circuit $U$ to the initial state $\ket{\psi_0}$ as follows:
\begin{align}
    \ket{\psi}=U\ket{\psi_0}. \label{eq:update_sv}
\end{align}
Here, we omit the parameter $\vec{\theta}$ for simplicity.
We assume that the quantum circuit $U$ is a product of $m$ quantum gates $U_0, U_1, \ldots, U_{m-1}$ arranged in an appropriate order. Consequently, Eq.~\eqref{eq:update_sv} can be rewritten as
\begin{align}
    \ket{\psi} =U_{m-1}\cdots U_1 U_0\ket{\psi_0},
    \label{eq:update_sv_gate}
\end{align}
Figure~\ref{fig:qsu_qct} provides an example of a quantum circuit for QSU, where $n=4$ and $m=5$.
In this example, $m=5$ gates are placed on the appropriate depth according to the data dependencies among the corresponding operators.
The depth of a gate here is the critical path length of the gate, and the circuit depth is the maximum depth in the circuit.

We now describe the concept of QSU with the QCT form.
The set $C$ of quantum operators consists of the quantum gates in Eq.~\eqref{eq:update_sv_gate}:
\begin{align}
    C =\{U_0,U_1,\ldots,U_{m-1}\}. \label{eq:operator_set_qsu}
\end{align}
As shown in Fig.~\ref{fig:qsu_qct}, the wires connected to gate $U_i$, where $0 \leq i<m$, represent the target qubit set $T(U_i)$ of the gate.

Given schedule $g$, QSU obtains the final state by
\begin{align}
    \ket{\psi}=g(m-1)\cdots g(1) g(0)\ket{\psi_0}.
    \label{eq:update_sv_reorder}
\end{align}
It is important to note that the schedule $g$ retains flexibility under constraint~C1 owing to the partially ordered nature of $C$.
That is, the schedule $g$ is allowed to be flexible, which facilitates solving our target problem, or the quantum operation scheduling.

\subsubsection{Expectation value computation}

EVC obtains a scalar value $\bra{\psi}H\ket{\psi}$ using the state vector $\ket{\psi}$.
The Hamiltonian $H$ can be expressed as the summation of $m$ Hamiltonian terms:
\begin{align}
    H=\sum_{i=0}^{m-1} a_i P_i, \label{eq:hamiltonian}
\end{align}
where $a_i$ and $P_i$ are the $i$-th real coefficient and the $i$-th Pauli string, respectively.
The Pauli string $P_i$ is a tensor product involving a combination with repetition of the identity matrix $I$ and the Pauli matrices $X$, $Y$, and $Z$ as shown in Eq.~\eqref{eq:pauli_form}.
\begin{align}
    P_i\in\{p_{n-1}\otimes p_{n-2}\otimes \cdots \otimes p_0 \mid p_j \in \{I,X,Y,Z\} ,0\leq j<n \}. \label{eq:pauli_form}
\end{align}
The definitions of the identity and Pauli matrices are given by
\begin{gather}
    I=\begin{bmatrix} 1 & 0 \\ 0 & 1 \end{bmatrix},
    X=\begin{bmatrix} 0 & 1 \\ 1 & 0 \end{bmatrix},
    Y=\begin{bmatrix} 0 & -i \\ i & 0 \end{bmatrix},
    Z=\begin{bmatrix} 1 & 0 \\ 0 & -1 \end{bmatrix}.
    \label{eq:pauli_matrix}
\end{gather}
Hence, EVC is defined by
\begin{align}
    \bra{\psi}H\ket{\psi}=\sum_{i=0}^{m-1} a_i \bra{\psi}P_i\ket{\psi}.
    \label{eq:evc}
\end{align}

\begin{figure}
    \centering
    \includegraphics[height=40mm]{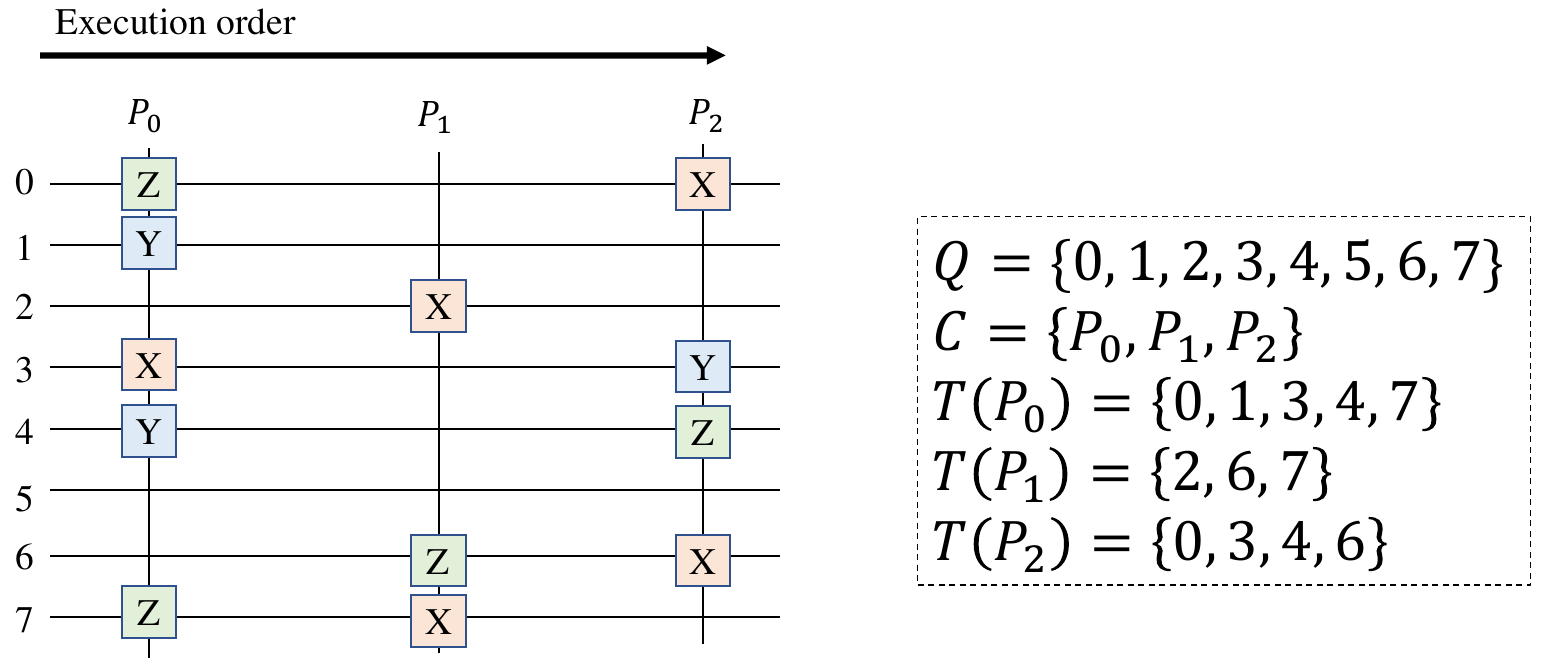}
    \caption{Diagram of EVC utilizing the QCT form. In this example, we assume that $P_0=Z_0 Y_1 X_3 Y_4 Z_7$, $P_1=X_2 Z_6 X_7$, and $P_2=X_0 Y_3 Z_4 X_6$.
    It is noteworthy that there is no data dependency among $P_0$, $P_1$, and $P_2$. Notice that this diagram, which is not a quantum circuit, illustrates a logical flow of EVC.}
    \Description[EVC logical flow with QCT form for an 8-qubit system.]{This diagram shows the logical flow of an Expectation Value Computation (EVC) on an 8-qubit system. It illustrates the execution order for three Pauli strings (P0, P1, P2), with gates representing which qubits each string acts upon. The box on the right gives the corresponding QCT representation.}
    \label{fig:evc}
\end{figure}

Similar to QSU, EVC can be described with the QCT form.
Figure~\ref{fig:evc} shows an example diagram of EVC.
In the case of VQE, EVC shares the same $Q$ with QSU.
$C$ corresponds to a set of the Pauli strings as follows:
\begin{align}
    C=\{P_0,P_1,\ldots,P_{m-1}\}. \label{eq:operator_set_evc}
\end{align}
$T(P_i)$ is the target qubit set on which the Pauli matrices comprising $P_i$ affect:
\begin{align}
    T(P_i)&=\{j \mid p_j\in \{X,Y,Z\}\}. \label{eq:pauli_target}
\end{align}
In Fig.~\ref{fig:evc}, we have $T(P_1)=\{2,6,7\}$ for $P_1=X_2 Z_6 X_7$, where $X_j$ ($Z_j$) is an $X$ ($Z$) operator applied to qubit $j$.

In the description mentioned above, $C$ forms an unordered set because the summation in Eq.~\eqref{eq:evc} is commutative.
As compared with the QSU, which holds partially ordered nature, the schedule $g$ in EVC is entirely flexible.

\subsection{Parallel state-vector-based quantum-circuit simulation}
p-SVQCS~\cite{p-svqcs} realizes parallel QSU on multiple processing elements (PEs) in a data-parallel manner.
Initially, the simulator divides and distributes the state vector across PEs.
The simulator iteratively performs multiplication of the distributed vector with gate matrices according to Eq.~\eqref{eq:update_sv_reorder}.
Between iterations, p-SVQCS conducts qubit reordering as necessary to update the distribution of the state vector.
The necessity of qubit reordering depends on the following factors: state vector distribution, qubit mapping, and the access pattern of the subsequent operation.

\begin{figure}
    \centering
    \includegraphics[height=55mm]{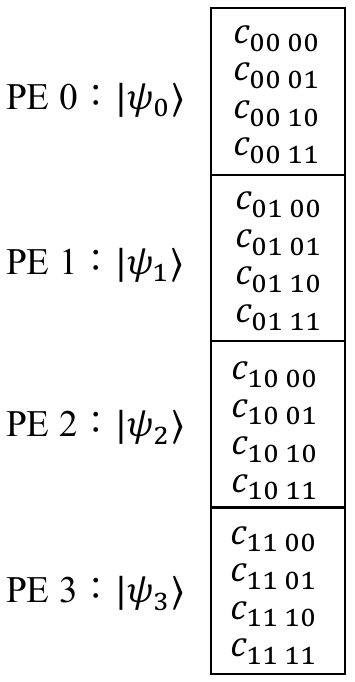}
    \caption{Example of state vector distribution in the case of a 4-qubit system on 4 PEs.
    A state vector for a 4-qubit system encompasses $2^4$ probability amplitudes, ranging from $c_{0000}$ to $c_{1111}$.
    PE $j$ deals with a sub state-vector $\ket{\psi_j}$ consisting of $2^4/4$ elements in p-SVQCS.
    For example, PE 0 and PE 1 handle elements from $c_{0000}$ to $c_{0011}$ and those from $c_{0100}$ to $c_{0111}$, respectively.
    }
    \Description[State vector distribution across PEs for a 4-qubit system.]{This figure illustrates how the state vector for a 4-qubit system (from amplitude c0000 to c1111) is partitioned and distributed across four Processing Elements (PEs). The vector is shown as a single column divided into four blocks, with each PE (0 to 3) holding one block.}
    \label{fig:sv_layout}
\end{figure}

\subsubsection{State vector distribution}
\label{sec:data_layout}

p-SVQCS divides the state vector into sub vectors using block decomposition.
Figure~\ref{fig:sv_layout} illustrates the state vector distribution of a 4-qubit system across 4 PEs.
Let $p$ be the number of PEs, assumed to be a power of two owing to the constraints of our distribution scheme.
PE $j$, where $0\leq j < p$, handles a sub state-vector $\ket{\psi_j}$ consisting of $2^n/p$ elements.
A vertical arrangement of the sub state-vectors $\ket{\psi_j}$ forms the whole state-vector $\ket{\psi}$ as follows:
\begin{align}
    \ket{\psi}=
    \begin{bmatrix}
        \ket{\psi_0} \\ \ket{\psi_1} \\ \vdots \\ \ket{\psi_{p-1}}
    \end{bmatrix},~
    \ket{\psi_j}=
    \begin{bmatrix}
        c_{(j)_2 0\ldots 00} \\ c_{(j)_2 0\ldots 01} \\ \vdots \\ c_{(j)_2 1\ldots 11}
    \end{bmatrix},
    \label{eq:sub_sv}
\end{align}
where $(x)_2$ denotes the binary representation of $x$.

\subsubsection{Qubit mapping}
\label{sec:qubit_distribution}
\begin{figure}
    \centering
    \includegraphics[height=30mm]{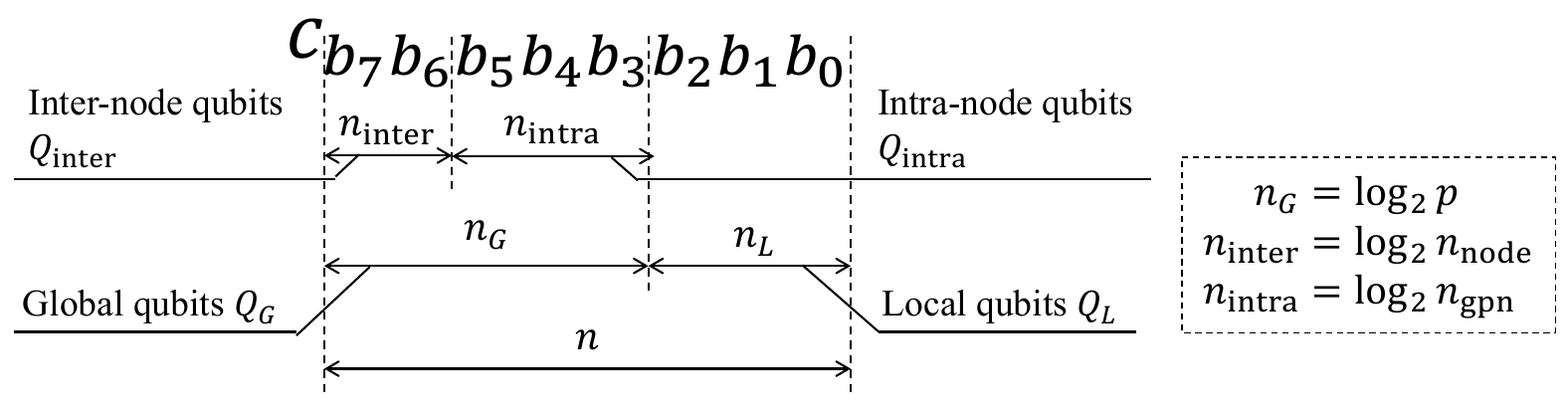}
    \caption{Qubit mapping scheme.
    This figure shows the binary index $(b_0, b_1, \ldots, b_7 \in \{0,1\})$ of the probability amplitude $c$ in an $n$-qubit system simulated on $n_\mathrm{node}$ nodes, each equipped with $n_\mathrm{gpn}$ GPUs.
    In this example, we set $n=8$, $n_\mathrm{node}=4$, and $n_\mathrm{gpn}=8$. Consequently, the total number of GPUs is $p=n_\mathrm{node}\times n_\mathrm{gpn}=32$.
    The dashed lines denote the qubit set divisions.
    The segmented sections of the binary index exhibit regularity based on the state-vector distribution across a two-layered interconnection.
    Within the sub state-vector on a GPU, the bits corresponding to local qubits are permuted from all zeros to all ones, while those corresponding to global qubits remain identical.
    Similarly, the bits corresponding to inter-node qubits remain identical across the sub state-vectors on a node.
    }
    \Description[Conceptual diagram of qubit mapping scheme.]{This is a conceptual diagram of a qubit mapping scheme. It shows how the binary index of a probability amplitude is segmented into local qubits (Q_L) and global qubits (Q_G). The global qubits are further subdivided into intra-node qubits (Q_intra) and inter-node qubits (Q_inter), corresponding to a hierarchical cluster architecture.}
    \label{fig:bit_part}
\end{figure}

Assuming block decomposition, we partition $Q$ into local and global divisions.
Let $Q_G$ and $Q_L$ denote a set of global qubits and that of local qubits, respectively.
We then have
\begin{align}
Q = Q_G \cup Q_L,
\label{eq:qubit_division}
\end{align}
where $Q_G \cap Q_L = \emptyset$.
Figure~\ref{fig:bit_part} illustrates the qubit mapping on the binary index of the probability amplitude $c$.
The most significant $n_G$ bits, where
$n_G=\log p$, are designated as global qubits~\cite{mpi-qulacs} because they determine PE $j$ that possesses $\ket{\psi_j}$, as depicted in Eq.~\eqref{eq:sub_sv}.
The remaining bits are termed local qubits, representing the local offset of the probability amplitude $c$ on PE $j$.

We further expand the concept of qubit mapping for a cluster of nodes, where each node incorporates multiple GPUs.
The interconnection of GPUs forms a two-layered topology: inter-node and intra-node connections.
By regarding each GPU as a PE, we categorize $Q_G$ into inter-node qubits $Q_\mathrm{inter}$ and intra-node qubits $Q_\mathrm{intra}$ as depicted in Fig.~\ref{fig:bit_part}.
Notations $n_\mathrm{node}$ and $n_\mathrm{gpn}$, which are both powers of two, denote the number of nodes and GPUs per node, respectively; thus, $p = n_\mathrm{node}\times n_\mathrm{gpn}$.
Given that block decomposition assigns PE $j$ to GPU $i$ on node $k$, we obtain
\begin{align}
    \ket{\psi_{k n_\mathrm{gpn} + i}}&=
    \begin{bmatrix}
         c_{(k)_2(i)_2 0\ldots00}\\ c_{(k)_2(i)_2 0\ldots01}\\ \vdots \\ c_{(k)_2(i)_2 1\ldots11}
    \end{bmatrix},
    \label{eq:sub_sv_gpu}
\end{align}
where $0 \leq i < n_\mathrm{gpn}$ and $0 \leq k < n_\mathrm{node}$.

\subsubsection{Access patterns of quantum operations}
\label{sec:access_pattern}

\begin{figure}[t]
  \begin{minipage}[b]{\linewidth}
    \centering
    \includegraphics[height=20mm]{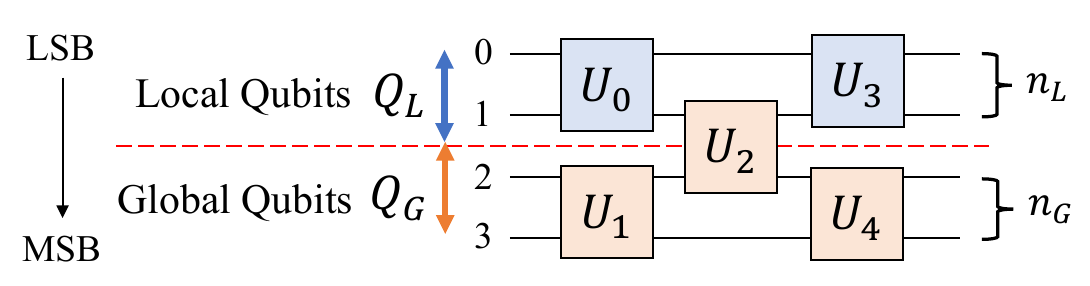}
    \subcaption{}
    \label{fig:qc_dep}
  \end{minipage}
  \\
  \begin{minipage}[b]{0.49\linewidth}
    \centering
    \includegraphics[height=55mm]{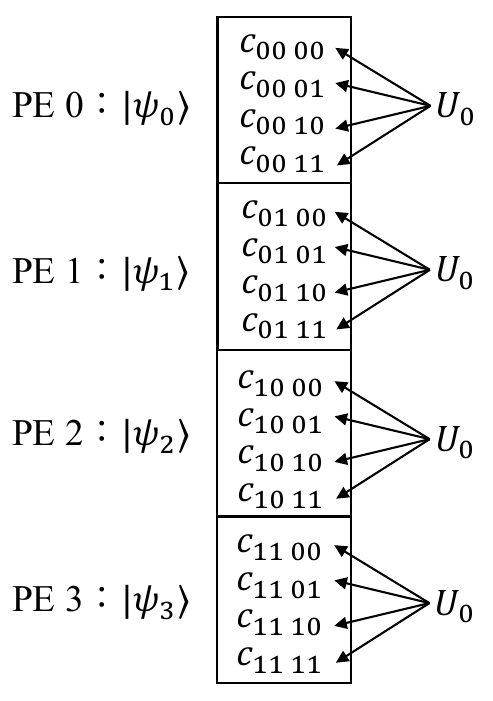}
    \subcaption{}
    \label{fig:narrow_dep}
  \end{minipage}
  \begin{minipage}[b]{0.49\linewidth}
    \centering
    \includegraphics[height=55mm]{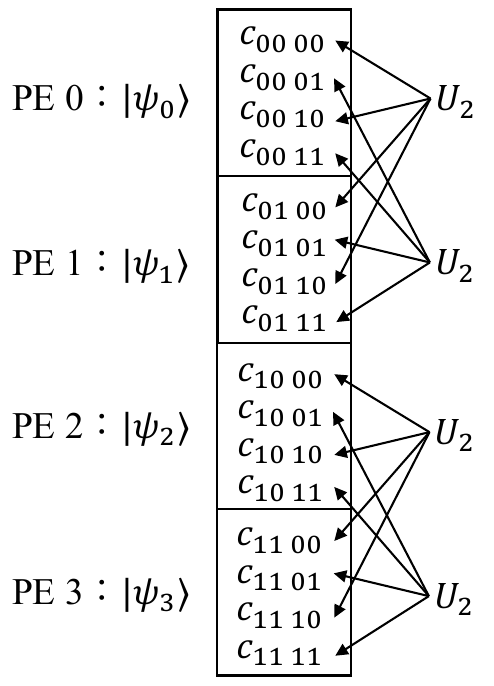}
    \subcaption{}
    \label{fig:wide_dep}
  \end{minipage}
  \caption{Examples of data-access patterns in a 4-qubit system simulated across four PEs.
    (a) The target quantum circuit and its qubit mapping. Blue gates $U_0$ and $U_3$ exhibit a narrow access pattern because $T(U_0) = T(U_3) = Q_L = \{0, 1\}$.
    In contrast, red gates $U_1$, $U_2$, and $U_4$ demonstrate a wide access pattern.
    (b) Narrow access. The arrows represent the input data required for updating $\ket{\psi_j}$ with gate $U_0$. For instance, PE 0 requires $c_{0000}$, \ldots, $c_{0011}$ when applying $U_0$ to $\ket{\psi}$.
    (c) Wide access. The four arrows pointing to $\ket{\psi_0}$ indicate that (1) PE 0 requires remote $c_{0100}$ and $c_{0110}$ from PE 1 to update $\ket{\psi_0}$ with $U_2$ and (2) PE 1 requires remote $c_{0001}$ and $c_{0011}$ from PE 0 to update $\ket{\psi_1}$ with $U_2$.}
    \Description[Data-access patterns (narrow vs. wide) in a 4-qubit, 4-PE simulation.]{This figure contains three parts. (a) shows a quantum circuit with local qubits {0,1} and global qubits {2,3}. (b) illustrates a "narrow access" pattern for gate U0, where each PE can compute using only its own data. (c) illustrates a "wide access" pattern for gate U2, where arrows indicate that data communication between PEs is required.}
  \label{fig:dep}
\end{figure}

Data-access patterns of operations depend on the qubit mapping.
We categorize access patterns into two types: \emph{narrow} and \emph{wide} access.
Figure~\ref{fig:dep} shows examples of these two access patterns.
Hereafter, we assume that operation $C_i$ corresponds to gate $U_i$, where $0 \leq i < m$.
Operation $C_i$ exhibits narrow access if its targets are within the local qubits, \textit{i.e.}, $T(C_i) \subseteq Q_L$; the operation demonstrates wide access, otherwise.
For narrow access, PE $j$ is allowed to locally compute the next values of $\ket{\psi_j}$ because the access pattern is confined within PE $j$.
Conversely, for wide access, PE $j$ requires remote data access across PEs to update $\ket{\psi_j}$, resulting in communication among PEs.

To reduce the communication overhead, the simulator determines a qubit mapping for each operation to ensure narrow access patterns.
Let $n_L$ be the number of local qubits in the circuit.
The possible set $Q_\mathrm{locals}$ of $n_L$ local qubits is then given by:
\begin{align}
    Q_\mathrm{locals}&=\{Q_L\in 2^Q \mid |Q_L|=n_L\}. \label{eq:local_qubits}
\end{align}
Let $M: C\rightarrow Q_\mathrm{locals}$ be a qubit map that realizes narrow accesses for the operations.
The map $M$ is constrained by
\begin{gather}
   \forall C_i\in C\,[T(C_i)\subseteq M(C_i)],
   \label{eq:qr_map_cond}
\end{gather}
where $0 \leq i<m$, which ensures narrow access for any operation in $C$. Hereafter, we call the above constraint as constraint C2.

\subsubsection{Parallel gate application}
PEs can simulate a gate application in a straightforward data-parallel manner when the gate exhibits a narrow access pattern.
Specifically, the simulator decomposes the matrix-vector product for a gate application into $p$ independent products; $U_i\ket{\psi}$, where $0 \leq i <m$, transforms into a tensor product of $n_G$ identity matrices and an $n_L$-qubit gate $U_i^\prime$:
\begin{align}
    U_i\ket{\psi}&=I^{\otimes n_G} \otimes U_i^\prime\ket{\psi} \notag \\
    &=\begin{bmatrix}
          U_i^\prime &  & 0 \\ & \ddots& \\ 0& & U_i^\prime
      \end{bmatrix}\begin{bmatrix}
          \ket{\psi_0} \\ \vdots \\ \ket{\psi_{p-1}}
      \end{bmatrix} \notag \\
    &=\begin{bmatrix}
          U_i^\prime\ket{\psi_0} \\ \vdots \\ U_i^\prime\ket{\psi_{p-1}}
      \end{bmatrix}. \label{eq:apply_parallel}
\end{align}
As a result, PE $j$, where $0 \leq j <p$, is allowed to simultaneously compute $U_i^\prime\ket{\psi_j}$ without communication because the data access of the product is closed in each PE.

Note that $U_i$ and $U_i^\prime$ are sparse matrices in general quantum circuits.
Hence, p-SVQCS ideally accelerates QSU by a factor of $p$ using $p$ PEs; the scale of the product is reduced by $1/p$, transitioning from the whole state vector $\ket{\psi}$ to a sub state-vector $\ket{\psi_j}$.

\subsubsection{Qubit reordering}
\label{sec:qr}
\begin{figure}[t]
  \begin{minipage}[t]{0.49\linewidth}
    \centering
    \includegraphics[width=0.7\linewidth]{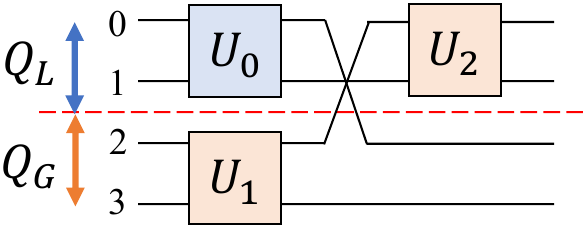}
    \subcaption{}
    \label{fig:qr_circuit}
  \end{minipage}
  \begin{minipage}[t]{0.49\linewidth}
    \centering
    \includegraphics[height=40mm]{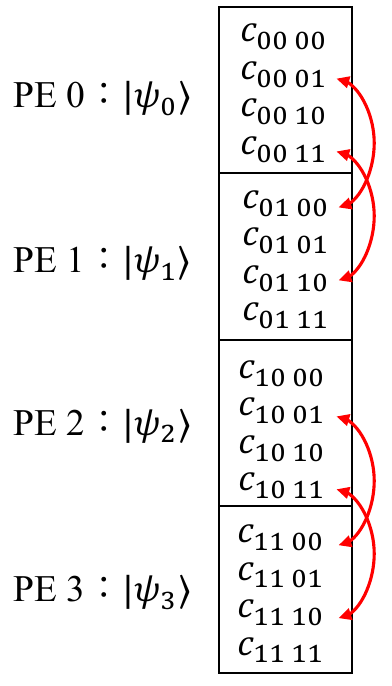}
    \subcaption{}
    \label{fig:qr_sv}
  \end{minipage}
    \caption{Example of qubit reordering in a 4-qubit system simulated accross four PEs. (a) Diagram illustrating qubit reordering, exchanging qubits $0 \in Q_L$ with $2 \in Q_G$.
    Qubit reordering entails the exchange of wires form the quantum circuit perspective.
    (b) Exchange of elements using qubit reordering. From the standpoint of state-vector distribution, qubit reordering involves the exchange in elements such as $c_{0001}$ and $c_{0100}$.
    As a result, the access pattern of quantum gate $U_2$ transitions from wide access to narrow access.}
    \Description[Qubit reordering changing gate access patterns from wide to narrow.]{This figure has two parts. (a) shows the reordering of qubit 0 (local) and qubit 2 (global) as crossing wires on a circuit diagram. (b) shows the same operation from the state vector perspective, where elements like c0001 and c0100 are swapped, transforming a gate's access pattern from wide to narrow.}
  \label{fig:qr_example}
\end{figure}
\begin{figure}[t]
  \begin{minipage}[b]{\linewidth}
    \centering
    \includegraphics[height=23mm]{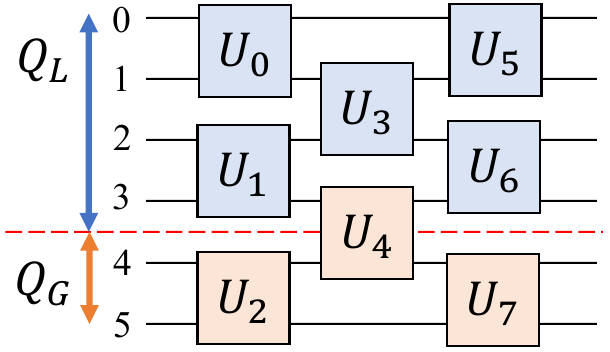}
    \subcaption{}
    \label{fig:qr_original}
  \end{minipage}
  \begin{minipage}[b]{0.55\linewidth}
    \centering
    \includegraphics[height=24mm]{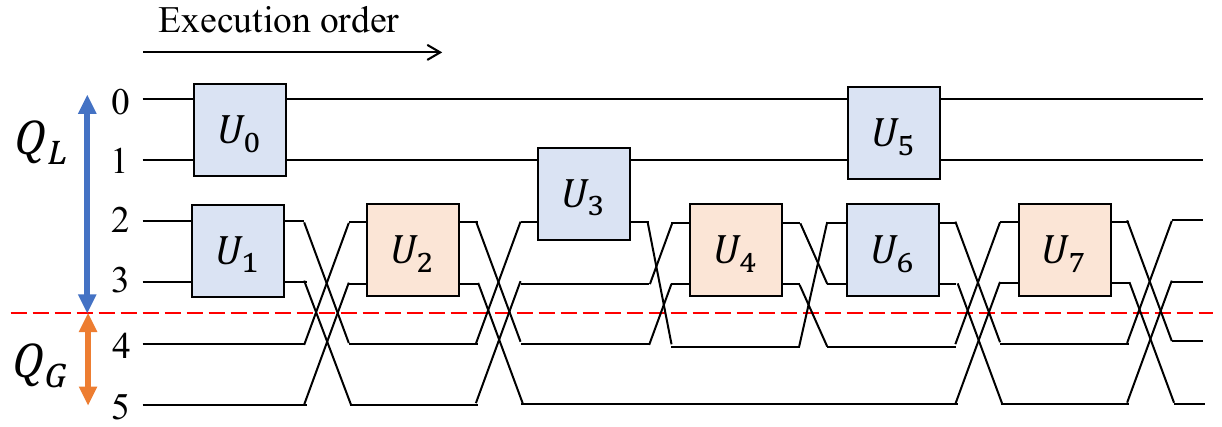}
    \subcaption{}
    \label{fig:qr_more}
  \end{minipage}
  \begin{minipage}[b]{0.44\linewidth}
    \centering
    \includegraphics[height=24mm]{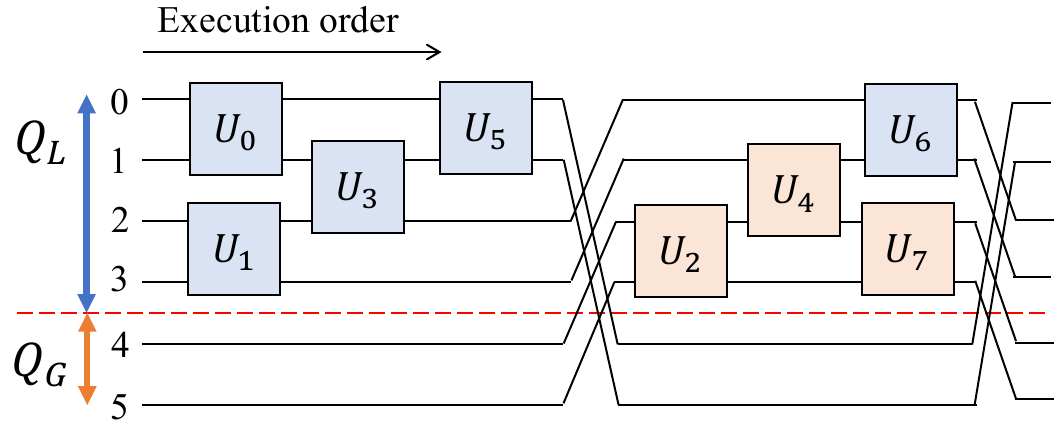}
    \subcaption{}
    \label{fig:qr_less}
  \end{minipage}
  \caption{Example of qubit reorderings that changes the number of qubit reorderings depending on the execution order.
    (a) The original quantum circuit to be simulated. The necessity for qubit reorderings arises because quantum gates $U_2, U_4,$ and $U_7$ have wide access.
    (b) A naive schedule organized in ascending order of gate depth: $g=(U_0, U_1, U_2, U_3, U_4, U_5, U_6, U_7)$. The simulator performs six qubit reorderings to complete the operation.
   (c) An efficient schedule wherein more gates successively exhibit narrow access within the same qubit mapping: $g=(U_0, U_1, U_3, U_5, U_2, U_4, U_6, U_7)$. The simulator replaces six qubit reorderings with two qubit reorderings by exchanging more qubits at once.
  }
  \Description[Impact of execution order on qubit reordering count.]{This figure shows how gate scheduling impacts efficiency. (a) is the original quantum circuit. (b) shows a naive, depth-first schedule that requires six qubit reorderings. (c) shows an efficient schedule that achieves the same result with only two reorderings by changing the gate order.}
  \label{fig:qr_defference}
\end{figure}

Qubit reordering is useful for ensuring narrow access by changing the qubit mapping.
In more detail, this technique exchanges a subset of $Q_L$ with an equal-sized subset of $Q_G$ as necessary before gate application.
Figure~\ref{fig:qr_example} shows an example of qubit reordering, which transforms wide accesses into narrow accesses by state vector redistribution.
When qubits $k$ and $l$ are exchanged, two elements whose indices differ only at the $k$-th and $l$-th bits are swapped, necessitating communication between PEs.
The necessity of qubit reordering depends on the current local qubits and the target of the next operation.
The simulator has to perform qubit reordering unless $T(g(i+1)) \subseteq M(g(i))$, where $0 \leq i<m$.
Figure~\ref{fig:qr_defference} depicts an example of qubit reorderings that reduces the number of qubit reorderings. This example explains that the necessity of qubit reordering relies on the schedule and qubit mapping.

Hence, the schedule $g$ and the qubit mapping $M$ determine the total number of qubit reorderings during the QSU process.
Equation~\eqref{eq:n_qr} defines the number of qubit reorderings in the QCT form.
\begin{align}
    N_\mathrm{qr}(Q,C,T,g,M)&=\sum_{i=0}^{|C|-1} \delta(M(g(i-1)), M(g(i))),\label{eq:n_qr}
\end{align}
where $M(g(-1))=\{0,1,\ldots,n_L-1\}$ is the initial set of local qubits, and the function $\delta$ is the Kronecker delta, which returns one if the two given arguments are different:
\begin{align}
    \delta(k,l)&=\begin{cases}
    1, & \mathrm{if~} k \neq l, \\
    0, & \mathrm{otherwise}.
\end{cases} \label{eq:f_eq}
\end{align}
p-SVQCS performs qubit reordering if $M(g(i-1))\neq M(g(i))$, \textit{i.e.}, if the succeeding operations have different qubit mappings.

\begin{figure}
    \centering
    \includegraphics[width=0.6\linewidth]{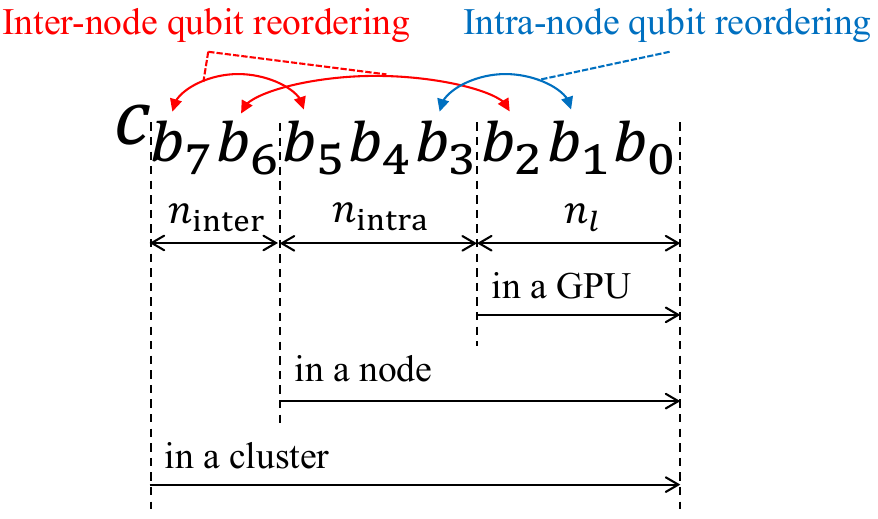}
    \caption{Diagram illustrating qubit reordering categorization. Intra-node qubit reorderings involve exchanging at least one qubit in $Q_\mathrm{intra}$ with one in $Q_L \cup Q_\mathrm{intra}$. Conversely, inter-node qubit reorderings entail exchanging a subset of $Q_\mathrm{inter}$ with one in the rest of $Q$.}
    \Description[Categorization of intra-node and inter-node qubit reordering.]{This diagram categorizes qubit reordering for a hierarchical cluster system. It distinguishes between "Intra-node qubit reordering" involving communication between GPUs on the same node, and "Inter-node qubit reordering" which involves more costly communication between different nodes.}
    \label{fig:qr_types}
\end{figure}

Furthermore, we characterize the overall communication cost of qubit reorderings within the QCT form. Assuming a cluster of multi-GPU nodes, we categorize qubit reorderings into two types: intra-node qubit reorderings and inter-node qubit reorderings, as illustrated in Fig.~\ref{fig:qr_types}.
Intra-node qubit reorderings involve exchanging state-vector elements among PEs on the same nodes, leading to intra-node communication.
Conversely, inter-node qubit reorderings entail inter-node communication, incurring higher overhead compared to intra-node qubit reorderings.
Equation~\eqref{eq:comm_cost} defines the total cost of qubit reorderings through the simulation.
\begin{align}
    \mathrm{Cost}(Q,C,T,g,M)&=w_\mathrm{intra}\times N_\mathrm{intra}(Q,C,T,g,M) + w_\mathrm{inter}\times N_\mathrm{inter}(Q,C,T,g,M)\label{eq:comm_cost},
\end{align}
where $N_\mathrm{intra}$ and $N_\mathrm{inter}$ denote the number of intra-node qubit reorderings and that of inter-node qubit reorderings, respectively, and $w_\mathrm{intra}$ and $w_\mathrm{inter}$ denote the weights associated with intra-node and inter-node communication, respectively.
These weights are determined by factors such as the latency and bandwidth of interconnections like NVLink and InfiniBand.
We assume that $w_\mathrm{inter}$ is tens of times larger than $w_\mathrm{intra}$.

\subsection{Problem definition}
\label{sec:problem_definition}
This study aims to find a solution $(g,M)$ that minimizes the total cost of qubit reorderings in both QSU and EVC.
Equation~\eqref{eq:opt_min} shows the common objective function in the QCT form.
\begin{equation}
\begin{aligned}
    \underset{g, M}{\mathrm{minimize}} \quad & \mathrm{Cost}(Q,C,T,g,M) & \label{eq:opt_min} \\
		\mathrm{subject~to} & \mathrm{~constraints~C1~and~C2~,~i.e.,~Eqs.}~(\ref{eq:gate_order_constraint}) \mathrm{~and~} (\ref{eq:qr_map_cond})\mathrm{,~respectively}.
\end{aligned}
\end{equation}
Constraint~C1 ensures maintaining data dependency, while constraint~C2 ensures narrow access.
In EVC, constraint~C1 always holds because the execution order of operations is completely flexible.

\section{Proposed Methods}
The proposed methods, which aim at reducing the total cost of qubit reorderings, are based on a lazy scheduling approach that processes operations with wide access as late as possible.
In other words, this lazy approach intensionally delays qubit reorderings because operations with wide access involve qubit reorderings.
By doing this, our approach processes more operations that can be processed with narrow access, increasing the number of operations per qubit reordering with less qubit reordering frequency.

The concept of the proposed methods is inspired by time-space tiling~\cite{time_space_tiling}, which processes operations tile by tile.
In the following, a tile represents a set of quantum gates that can be processed with narrow access under a qubit mapping.
We resolve the qubit reordering-cost minimization problem (Eq.~(\ref{eq:opt_min})) as a tiling problem that covers the quantum circuit with the minimum number $t$ of tiles.
Given $Q_L\in Q_\mathrm{locals}$, a tile $G$ satisfies the following condition:
\begin{align}
    G \subseteq \{C_i\in C \mid T(C_i)\subseteq Q_L \}. \label{eq:tile_condition}
\end{align}
The simulation requires $(t-1)$ qubit reorderings for $t$ tiles because a tile transition corresponds to a qubit reordering.
Reducing the number $t$ of tiles increases the tile size, leading to lazy qubit reorderings.

We introduce two tiling-based methods to suppress the occurrence of qubit reorderings in QSU and EVC, providing a solution $(g, M)$ for Eq.~(\ref{eq:opt_min}).
The first method empirically yields a high-quality solution for QSU in polynomial time by iteratively selecting the locally optimal qubit mapping.
Meanwhile, the second method, designed for EVC, offers an optimal solution for simulations of feasible scale by leveraging the arbitrary order of operations.

\subsection{Time-space tiling for quantum state update}

\begin{figure}
    \centering
    \includegraphics[width=.7\linewidth]{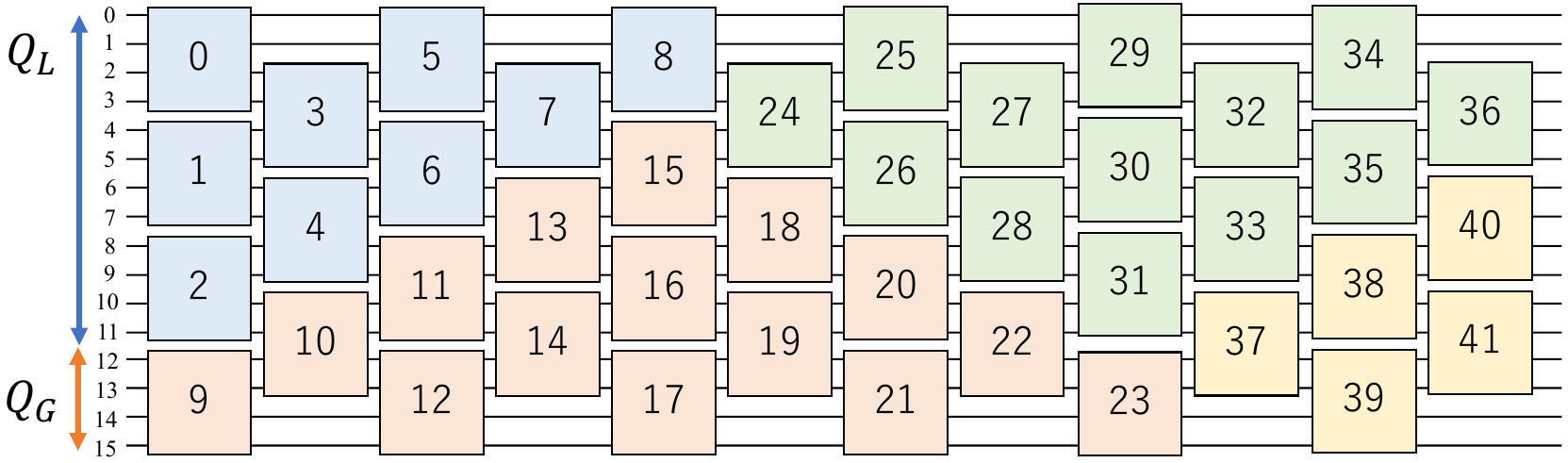}
    \caption{Flat tiling for a QSU with the GateFabric circuit.
    The numbers on the quantum gates denote their application order.
    QSU is performed on $16$ GPUs.
    Each group of quantum gates sharing the same color forms a tile.
    These tiles indicate that p-SVQCS applies quantum gates with identical qubit mappings.
    Consequently, p-SVQCS requires a qubit reordering if the $i$-th and $(i+1)$-th quantum gates have different colors.
    For instance, p-SVQCS applies the 0-th and 8-th quantum gates with the same qubit mapping.
    Conversely, the 8-th and 9-th quantum gates entail distinct qubit mappings.
    Subsequently, qubit reordering adjusts the global qubits from $Q_G=\{12,13,14,15\}$ to $Q_G=\{0,1,2,3\}$.
    During the transition from the 23rd to the 24th gate, qubit reordering changes to the original global qubits $Q_G=\{0,1,2,3\}$.
    Then, p-SVQCS repeats a similar process and results in 3 qubit reorderings in this QSU scenario.
    }
    \Description[Flat tiling applied to a GateFabric quantum circuit.]{This figure shows the application of the flat tiling method to a GateFabric circuit simulation. Quantum gates of the same color form a "tile" and are executed with the same qubit mapping. A change in color indicates that a qubit reordering is performed. In this example, the entire simulation completes with three reorderings.}
    \label{fig:flat_tiling}
\end{figure}

The proposed methods utilize time-space tiling for quantum circuits, where the gate depth and qubits correspond to time and space, respectively.
Figure~\ref{fig:flat_tiling} shows how time-space tiling can be utilized for the GateFabric circuit~\cite{gatefabric}, or a variational quantum circuit used in VQE.
This technique locally maximizes the number of quantum gates applied per qubit reordering by rearranging the application order of gates.
While the tiling method for QSU is heuristic, it provides a high-quality solution leveraging the principles of time-space tiling and the dependencies among quantum gates.

We introduce two tiling methods tailored to the interconnection topology of GPU clusters.
\begin{enumerate}
    \item Flat tiling: designed for flat network topologies commonly found in general cases, such as the star topology.
	\item Hierarchical tiling: developed to accommodate hierarchical network topologies, building upon the principles of flat tiling.
\end{enumerate}
Flat tiling reduces the total number of qubit reorderings, while hierarchical tiling preferentially reduces the occurrence of inter-node qubit reorderings over intra-node qubit reorderings.
In scenarios where the interconnection of a cluster exhibits a hierarchical structure, hierarchical tiling outperforms flat tiling, offering a more effective solution.

\begin{algorithm}[t]
\caption{Flat tiling for QSU.}
\label{alg:flat_tiling}
\KwIn{set $Q=\{0,1,\ldots,n-1\}$ of qubits, ordered set $C=\{U_0,U_1,\ldots,U_{m-1}\}$ of quantum gates, target map $T:C\rightarrow 2^Q$, and number $n_L$ of local qubits.}
\KwOut{schedule $g:\{0,1,\ldots,m-1\}\rightarrow C$ and qubit mapping $M:C\rightarrow Q_\mathrm{locals}$.}
\SetKwProg{function}{Function \algoname{FlatTimeSpaceTiling}}{}{end}

\function{$(Q,C,T,n_L)$}{
        $Q_L \gets \{0,1,\ldots,n_L-1\}$ \Comment*[r]{Initial local qubits}
        $C_\mathrm{remain} \gets ($Sort $C$ in ascending order of depths$)$ \Comment*[r]{Satisfy constraint C1}
        $x \gets 0$\;
        \While {$C_\mathrm{remain} \neq \emptyset$}{
            $G\gets$ \algoname{TileConstruction}$(Q_L,C_\mathrm{remain},T)$ \Comment*[r]{Process (P1) returns a tile}
            \For(\Comment*[f]{For every gate in the tile}){$j=0 \ldots |G|-1$}{
                $g(x)\gets j$-th gate of tile $G$\;
                $M(g(x))\gets Q_L$ \Comment*[r]{Assign the local qubits to the gate}
                $x\gets x+1$\;
                $C_\mathrm{remain}\gets C_\mathrm{remain} \setminus \{g(x)\}$\;
            }
            $Q_L \gets$ \algoname{QubitMapping}$(Q,C_\mathrm{remain},T,n_L)$ \Comment*[r]{Process (P2) returns the updated local qubits}
        }
        \Return $(g,M)$\;
}
\end{algorithm}

\begin{figure}[t]
  \begin{minipage}[b]{0.32\linewidth}
    \centering
    \includegraphics[width=0.9\linewidth]{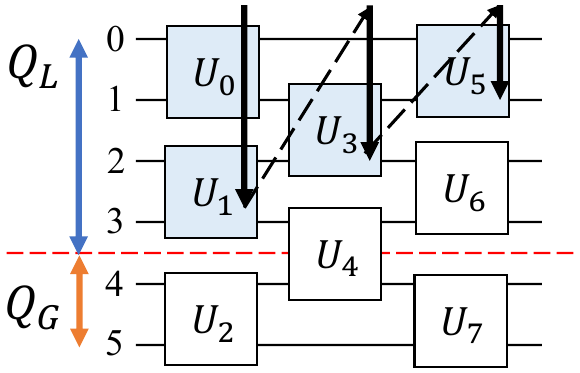}
    \subcaption{}
    \label{fig:tiling_s1_p1}
  \end{minipage}
  \begin{minipage}[b]{0.32\linewidth}
    \centering
    \includegraphics[width=0.9\linewidth]{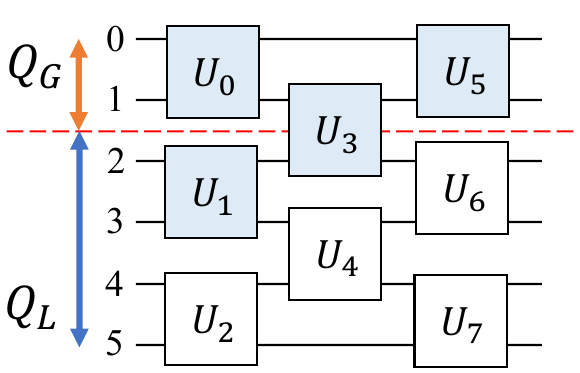}
    \subcaption{}
    \label{fig:tiling_s1_p2}
  \end{minipage}
  \begin{minipage}[b]{0.32\linewidth}
    \centering
    \includegraphics[width=0.9\linewidth]{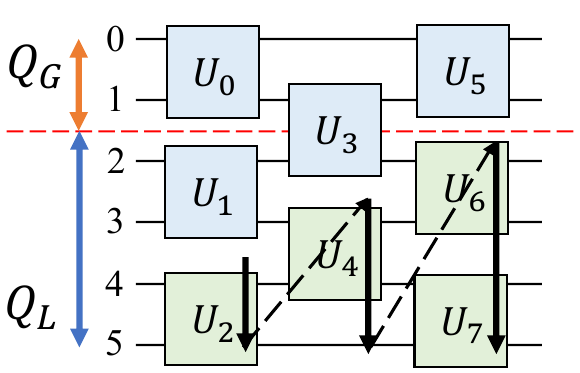}
    \subcaption{}
    \label{fig:tiling_s2_p1}
  \end{minipage}
  \caption{An overview of the tiling algorithm for a quantum circuit.
    (a) First, (P1) localizes quantum gates $U_0$, $U_1$, $U_3$, and $U_5$ with respect to $Q_L$ and arranges them in depth-first order.
    (b) Next, (P2) updates the qubit mapping so that qubits 2, 3, 4, and 5, affected by the remaining quantum gates $U_2$, $U_4$, and $U_6$, are selected as local qubits.
    (c) Finally, (P1) localizes and orders the quantum gates $U_2$, $U_4$, $U_6$, and $U_7$ according to the new qubit mapping.
    The tiling process terminates when all quantum gates have been ordered.
    As a result, time-space tiling generates a single qubit reordering for this circuit.}
    \Description[Overview of the quantum circuit tiling algorithm.]{This three-panel figure illustrates how the tiling algorithm works. (a) The algorithm identifies and groups executable gates into a first tile based on the current mapping. (b) It then updates the qubit mapping (a reordering) to suit the remaining gates. (c) Finally, it groups the remaining gates into a new tile under the new mapping.}
  \label{fig:time_space_tiling_overview}
\end{figure}

\begin{algorithm}[t]
\caption{Tile construction and gate scheduling.}
\label{alg:construct_tile}
\KwIn{set $Q_L$ of local qubits, ordered set $C$ of quantum gates, and target map $T$.}
\KwOut{Tile $G$ of ordered quantum gates.}
\SetKwProg{function}{Function \algoname{TileConstruction}}{}{end}

\function{$(Q_L,C,T)$}{
        $Q_\mathrm{avail} \gets Q_L$ \Comment*[r]{Available local qubits}
        $G\gets \emptyset$\;
        $C^\prime\gets C$ \Comment*[r]{Gates in $C^\prime$ are ordered by depth}
        \For(\Comment*[f]{For every gate in $C^\prime$}){$i=0\ldots |C^\prime|-1$}{
            $U \gets$ $i$-th gate of $C^\prime$\;
            \uIf(\Comment*[f]{Satisfy constraint~C2}){$T(U) \subseteq Q_\mathrm{avail}$}{
                Append $U$ to the tail of $G$\;
                $C^\prime \gets C^\prime\setminus \{U\}$\;
            }\Else{
                $Q_\mathrm{avail} \gets Q_\mathrm{avail} \setminus T(U)$ \Comment*[r]{Qubits blocked by gate $U$ are no longer available}
            }
            \If {$Q_\mathrm{avail} = \emptyset$}{
                \textbf{break}
            }
        }
        \Return $G$
}
\end{algorithm}

\begin{algorithm}[t]
\caption{Qubit mapping based on greedy selection of local qubits.}
\label{alg:update_qubit_division}
\KwIn{set $Q$ of qubits, ordered set $C$ of quantum gates, target map $T$, and number $n_L$ of local qubits.}
\KwOut{Updated set $Q_L$ of local qubits.}
\SetKwProg{function}{Function \algoname{QubitMapping}}{}{end}

\function{$(Q,C,T,n_L)$}{
        $Q_\mathrm{avail} \gets Q$\;
        $Q_L\gets \emptyset$\;
        $C^\prime\gets C$ \Comment*[r]{Gates in $C^\prime$ are ordered by depth}
        \For(\Comment*[f]{For every gate in $C^\prime$}){$i=0\ldots |C^\prime|-1$}{
            $U \gets$ $i$-th gate of $C^\prime$\;
            \uIf(\Comment*[f]{Satisfy constraint C2}){$T(U) \subseteq Q_\mathrm{avail} \land |Q_L\cup T(U)|\leq n_L$}{
                $Q_L\gets Q_L \cup T(U)$
            }\Else{
                $Q_\mathrm{avail} \gets Q_\mathrm{avail} \setminus T(U)$
            }
            \If {$Q_\mathrm{avail} = \emptyset$}{
                \textbf{break}
            }
        }
        \If(\Comment*[f]{Only used in hierarchical tiling}){$Q_L=\emptyset$}{
            \Return $\emptyset$
        }
        \While(\Comment*[f]{Boundary treatment for the last tile}){$|Q_L| < n_L$}{
            $q\gets$ the lowest qubit in $(Q \setminus Q_L)$\;
            $Q_L\gets Q_L\cup\{q\}$\;
        }
        \Return $Q_L$
}
\end{algorithm}

\subsubsection{Flat tiling}
\label{sec:flat_tiling}

Flat tiling alternates the following two processes iteratively until all the quantum gates are ordered.
\begin{description}
	\item[(P1)] Constructing a local maximum tile under a given local qubits.
	\item[(P2)] Updating the qubit mapping for the remaining quantum gates.
\end{description}
We present the full algorithm in Algorithm~\ref{alg:flat_tiling}.
At the beginning (line 3), the proposed methods sort all gates by their depth to satisfy constraint~C1.
For gates at the same depth, the sorting algorithm prioritizes the target qubits in ascending order so that the gates are adjacent to each other.
Algorithms~\ref{alg:construct_tile} and \ref{alg:update_qubit_division} show the details of (P1) and (P2), respectively.

First, (P1) constructs a tile by localizing quantum gates that are executable with the given local qubits $Q_L$.
Because simulation is performed on a per-tile basis, (P1) excludes gates that indirectly depend on a global qubit.
For example, $U_6$ in Fig.~\ref{fig:tiling_s1_p1} is excluded from the tile being constructed because $U_6$ depends on global qubit 4 through $U_4$.
In other words, the target qubits $T(U)$ of an inexecutable gate $U$ cannot be available for the following gates in the tile.
After that, Algorithm~\ref{alg:construct_tile} updates available qubits by removing unavailable qubits from $Q_\mathrm{avail}$ at line~11.
Note that condition $T(U) \subseteq Q_{\mathrm{avail}}$ at line~7 is equivalent to constraints~C1 and C2 because the proposed method treats gates in depth order.

Second, (P2) updates the qubit mapping, which greedily selects local qubits such that the remaining gates at shallow depths are preferentially processed at the next iteration.
Such shallow gates tend to have more dependencies on deeper gates.
This implies that the target qubits of shallow gates must be selected as local qubits to maximize the number of available qubits (line 8 of Algorithm~\ref{alg:update_qubit_division}).
Thus, the proposed method greedily selects the target qubits of shallow gates for the next local qubits, which contributes to enlarge the current tile.

The time complexity of the flat tiling algorithm is $\mathcal{O}(\overline{d}m)$, where $\overline{d}$ is the average number of target qubits.
The time complexity can be analyzed as follows.
First, the sorting process at line~3 of Algorithm~\ref{alg:flat_tiling} can be processed with bucket sort in $\mathcal{O}(\overline{d}m)$ time.
In more detail, we regard a quantum bit as a bucket and insert every gate to the bucket that corresponds to the target qubit of the gate.
After iterating this insertion operation $m$ times, the sorting process can be completed by taking the first gate from the buckets.
The time complexity of this process is $\mathcal{O}(\overline{d}m)$ time because an insertion operation takes $\mathcal{O}(\overline{d})$ time.
This process is needed only once at the beginning of Algorithm~\ref{alg:flat_tiling} because sorted gates can be reused for succeeding processes.
Second, a single call of \algoname{TileConstruction} takes $\mathcal{O}(|G|+k)$ time because the dominant loop at line~5 of Algorithm~\ref{alg:construct_tile} iterates ($|G|+k$) times, where $|G|$ and $k$ correspond to the total number of gates appended at line 8 and that of (nonappended) gates processed at line 11, respectively.
That is, $|G|$ gates are processed with $\mathcal{O}($|G|$+k)$ time, and thereby the cumulative time complexity for \algoname{TileConstruction} is $\mathcal{O}(m)$ because the summation of tile sizes is $m$ and $\sum k \in \mathcal{O}(m)$.
Third, \algoname{QubitMapping} requires $\mathcal{O}(m)$ time by mapping all gates in $C$.
Thus, the overall time complexity with flat tiling is $\mathcal{O}(\overline{d}m)$.
Hence, the scheduling time of this method is negligible in the overall execution time for QSU, which is exponential in time for $n$.

\subsubsection{Hierarchical tiling}
\label{sec:hiera_tiling}

\begin{figure}
    \centering
    \includegraphics[width=.75\linewidth]{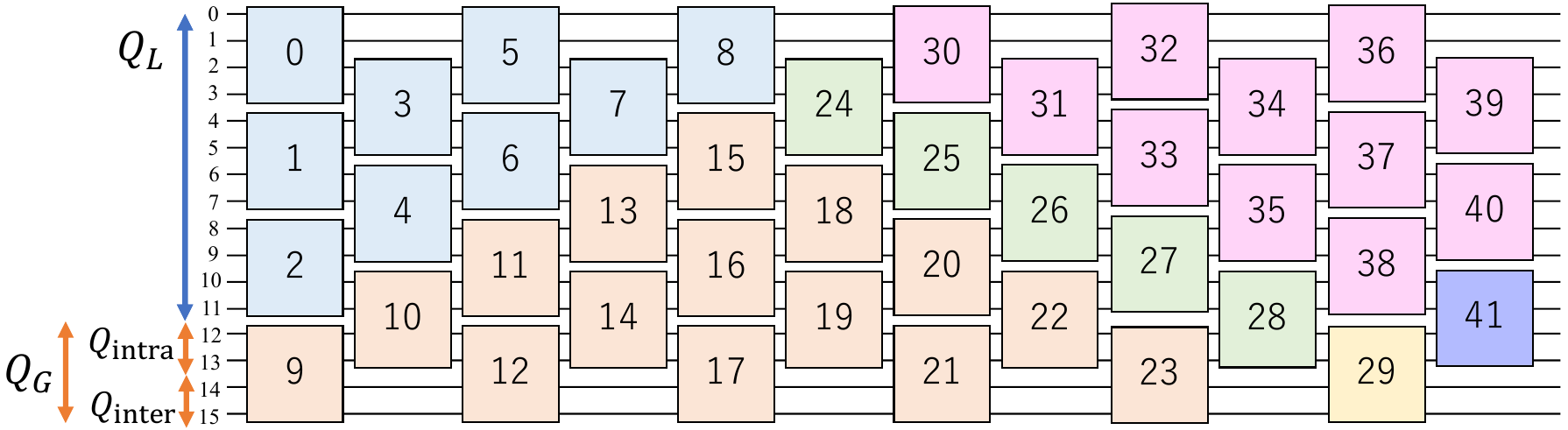}
    \caption{Hierarchical tiling for a QSU with the GateFabric circuit.
    This circuit is the same as in Fig.~\ref{fig:flat_tiling}.
    In this example, p-SVQCS executes two inter-node qubit reorderings and three intra-node qubit reorderings.
    First, during the transition from the 8th to the 9th gate, an inter-node qubit reordering changes the intra- and inter-node qubits from $Q_\mathrm{intra}=\{12,13\}$ and $Q_\mathrm{inter}=\{14,15\}$ to $Q_\mathrm{intra}=\{2,3\}$ and $Q_\mathrm{inter}=\{0,1\}$, respectively.
    Second, in the transition from the 23rd to the 24th gate, an intra-node qubit reordering modifies the intra-node qubits to $Q_\mathrm{intra}=\{14,15\}$ while maintaining the same inter-node qubits.
    Third, another intra-node qubit reordering adjusts the intra-node qubits to $Q_\mathrm{intra}=\{2,3\}$, which results in the 29th gate becoming applicable.
    Fourth, inter-node qubit reordering changes to initial inter- and intra-node qubits $Q_\mathrm{intra}=\{12,13\}$ and $Q_\mathrm{inter}=\{14,15\}$, respectively, in the transition from the 29th to the 30th gate.
    Finally, intra-node qubit reordering changes $Q_\mathrm{intra}=\{0,1\}$ to apply the 41st gate.}
    \Description[Hierarchical tiling applied to a GateFabric quantum circuit.]{This shows the hierarchical tiling method applied to the same circuit from Figure 9. The coloring illustrates that this method accepts more, lower-cost intra-node reorderings in order to minimize high-cost inter-node reorderings. The result is two inter-node and three intra-node reorderings.}
    \label{fig:hierarchical_tiling}
\end{figure}

\begin{algorithm}[t]
\caption{Hierarchical tiling for QSU.}
\label{alg:hierarchical_tiling}
\KwIn{set $Q=\{0,1,\ldots,n-1\}$ of qubits, ordered set $C=\{U_0,U_1,\ldots,U_{m-1}\}$ of quantum gates, target map $T:C\rightarrow 2^Q$, number $n_L$ of local qubits, and number $n_{\mathrm{intra}}$ of intra-node qubits.}
\KwOut{schedule $g:\{0,1,\ldots,m-1\}\rightarrow C$ and qubit mapping $M:C\rightarrow Q_\mathrm{locals}$.}
\SetKwProg{function}{Function \algoname{HierarchicalTimeSpaceTiling}}{}{end}

\function{$(Q,C,T,n_L,n_{\mathrm{intra}})$}{
        $Q_L \gets \{0,1,\ldots,n_L-1\}$ \Comment*[r]{Initial local qubits}
        $Q_{\mathrm{intra}} \gets \{n_L,n_L+1,\ldots,n_L+n_{\mathrm{intra}}-1\}$\;
        $C_\mathrm{remain} \gets ($Sort $C$ in ascending order of depths$)$ \Comment*[r]{Satisfy constraint C1}
        $x \gets 0$\;
        \While {$C_\mathrm{remain} \neq \emptyset$}{
            $G\gets$ \algoname{TileConstruction}{$(Q_L,C_\mathrm{remain},T)$} \Comment*[r]{Process (P1) returns a tile}
            \For(\Comment*[f]{For every gate in the tile}){$j=0\ldots |G|-1$}{
                $g(x)\gets j$-th gate of tile $G$\;
                $M(g(x))\gets Q_L$ \Comment*[r]{Assign the local qubits to the gate}
                $x\gets x+1$\;
                $C_\mathrm{remain}\gets C_\mathrm{remain} \setminus \{g(x)\}$\;
            }
            $(Q_L, Q_{\mathrm{intra}}) \gets$ \algoname{HierarchicalQubitMapping}{$(Q,C_\mathrm{remain},T,Q_L,Q_{\mathrm{intra}})$} \Comment*[r]{Process (P2)}
        }
        \Return $(g,M)$\;
}
\end{algorithm}

\begin{algorithm}[t]
\caption{Hierarchical qubit mapping based on greedy selection of local qubits.}
\label{alg:hierarchical_update_qubit_division}
\KwIn{set $Q=\{0,1,\ldots,n-1\}$ of qubits, ordered set $C=\{U_0,U_1,\ldots,U_{m-1}\}$ of quantum gates, target map $T:C\rightarrow 2^Q$, set $Q_L$ of local qubits, and set $Q_\mathrm{intra}$ of intra-node qubits.}
\KwOut{updated set $Q_L^\prime$ of local qubits and updated set $Q_\mathrm{intra}^\prime$ of intra-node qubits.}
\SetKwProg{function}{Function \algoname{HierarchicalQubitMapping}}{}{end}

\function{$(Q,C,T,Q_L,Q_\mathrm{intra})$}{
        $Q_L^\prime \gets$\algoname{QubitMapping}{$(Q_L\cup Q_\mathrm{intra},C,T,|Q_L|)$} \Comment*[r]{First phase for an intra-node qubit reordering}
        \If {$Q_L^\prime\neq \emptyset$}{
            $Q_\mathrm{intra}^\prime \gets (Q_L\cup Q_\mathrm{intra}) \setminus Q_L^\prime$
        }\Else{
            $Q_L^\prime \gets$\algoname{QubitMapping}{$(Q,C,T,|Q_L|)$} \Comment*[r]{Second phase for an inter-node qubit reordering}
            $Q^\prime \gets$\algoname{QubitMapping}{$(Q,C,T,|Q_L|+|Q_\mathrm{intra}|)$}\;
            $Q_\mathrm{intra}^\prime \gets Q^\prime\setminus Q_L^\prime$\;
        }
        \Return $(Q_L^\prime,Q_\mathrm{intra}^\prime)$\;
}
\end{algorithm}

This approach offers a solution to reduce the overall qubit reordering costs by accounting for the hierarchical interconnection of a GPU cluster.
As described in Section~\ref{sec:qr}, we assume a two-layered interconnection where inter-node connection is tens of times slower than intra-node connection.
In such a situation, it is worth reducing inter-node qubit reorderings even if intra-node qubit reorderings increase instead.
Therefore, the hierarchical tiling method accepts generating intra-node qubit reorderings to minimize the occurrence of inter-node qubit reorderings.
In Fig.~\ref{fig:hierarchical_tiling}, hierarchical tiling has two additional qubit reorderings compared to flat tiling (Fig.~\ref{fig:flat_tiling}).
However, hierarchical tiling reduces the total qubit reordering costs by replacing one inter-node qubit reordering with three intra-node qubit reorderings.

The hierarchical tiling method can be realized by extending the flat tiling method.
As shown in Algorithm~\ref{alg:hierarchical_tiling}, the algorithm differs from Algorithm~\ref{alg:flat_tiling} at lines 3 and 13, which initializes the set of intra-node qubits and replaces the call of \algoname{QubitMapping} with that of \algoname{HierarchicalQubitMapping} to take advantage of the two-layered interconnection, respectively.
A key strategy of \algoname{HierarchicalQubitMapping} is two-phased selection of qubit mapping (Algorithm~\ref{alg:hierarchical_update_qubit_division}).
In the first phase, hierarchical tiling tries to select $n_L$ qubits for local qubits from the current $Q_L \cup Q_{\mathrm{intra}}$ (line~2), which excludes inter-node qubits; flat tiling selects from all qubits.
If the first phase finds a valid qubit mapping, the qubit mapping always leads to an intra-node qubit reordering.
Otherwise, the second phase selects the local and intra-node qubits from all qubits (lines 6 and~7), resulting in an inter-node qubit reordering.
Thus, the hierarchical tiling method prioritizes intra-node qubit reorderings over inter-node qubit reorderings, which effectively reduces the total qubit reordering cost.

\subsection{Time-space tiling for expectation value computation}
\label{sec:proposed_evc}
We employ time-space tiling in EVC to utilize the same concept as QSU.
Our approach involves creating tiles comprising Pauli strings to execute qubit reorderings as lazily as possible, mirroring the strategy employed in QSU.
Our method effectively reduces the frequency of qubit reorderings by leveraging EVC's inherent flexibility in operation ordering.

The proposed EVC method consists of the following three components, which offer an optimal solution for minimizing qubit reordering costs.
\begin{enumerate}
    \item Parallel EVC with qubit reordering.
    \item Tiling Pauli strings to aggregate qubit reorderings.
    \item Diagonalization of Pauli strings to minimize of the number of qubit reorderings.
\end{enumerate}
The first component enhances the locality of qubit references to reduce communication.
Next, the second component forms tiles comprising Pauli strings to decrease the communication frequency for qubit reordering-based parallel EVC.
Finally, the third component transforms Pauli strings into efficient forms to reduce the number of qubit reordering occurences.

\subsubsection{Parallel EVC with qubit reordering}
\begin{figure}[t]
  \begin{minipage}[b]{0.49\linewidth}
    \centering
    \includegraphics[width=\linewidth]{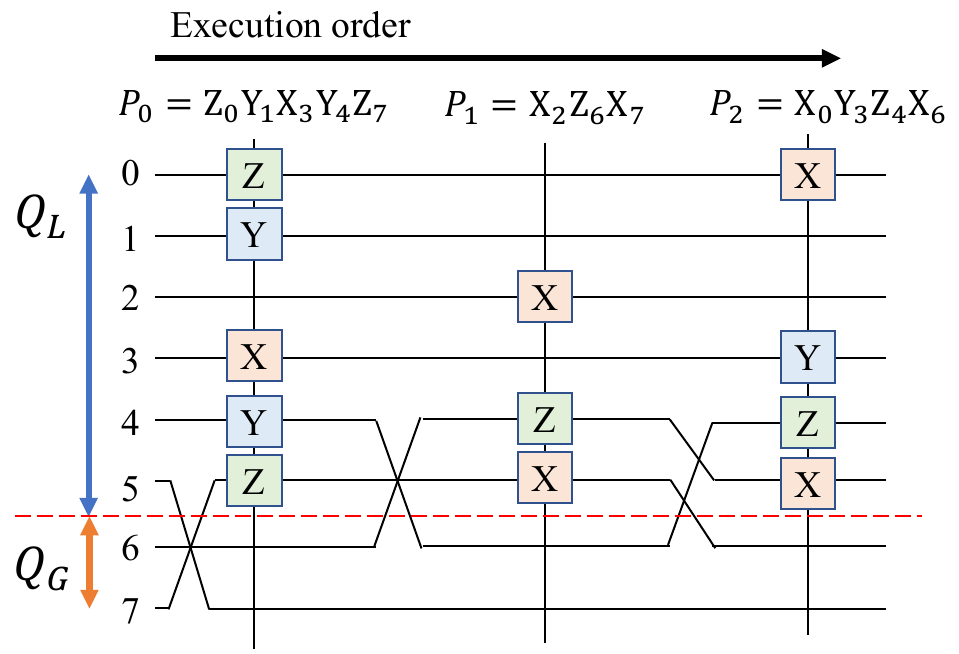}
    \subcaption{}
    \label{fig:parallel_evc_naive}
  \end{minipage}
  \begin{minipage}[b]{0.49\linewidth}
    \centering
    \includegraphics[width=\linewidth]{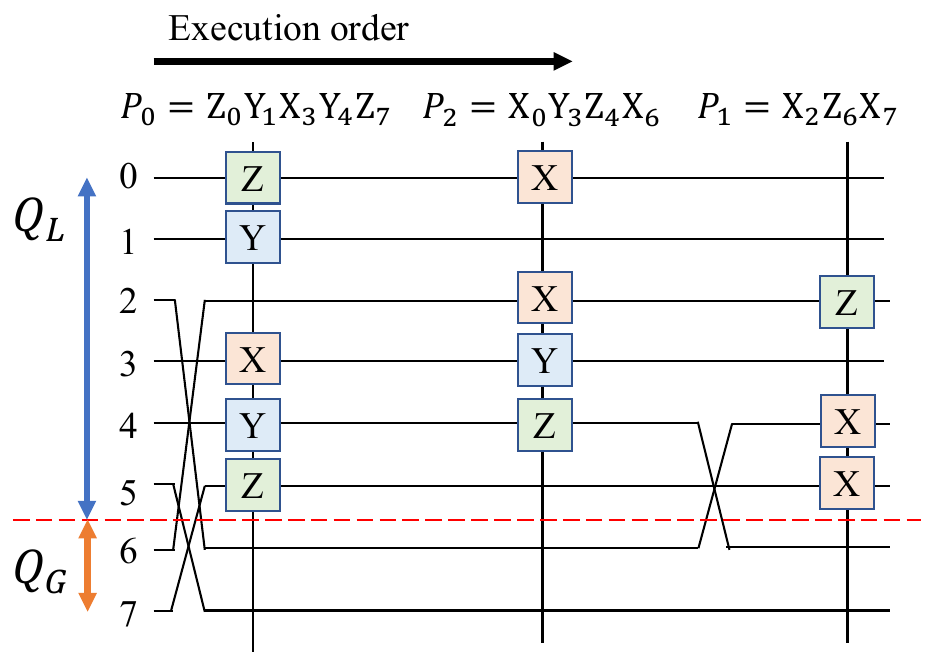}
    \subcaption{}
    \label{fig:parallel_evc_efficient}
  \end{minipage}
  \caption{Examples of qubit reordering-based parallel EVC for Fig.~\ref{fig:evc}.
  (a) Index order, which follows the order of Pauli string indices, requires three qubit reorderings.
  (b) Efficient order, which exchanges $P_1$ with $P_2$, reduces the requirement to two qubit reorderings.
  }
  \Description[Optimizing qubit reordering in parallel EVC.]{This figure demonstrates the benefit of reordering operations in EVC. (a) shows that computing Pauli strings in their default index order requires three reorderings. (b) shows that by changing the computation order, the number of required reorderings can be reduced to two.}
  \label{fig:parallel_evc}
\end{figure}

We propose an effective qubit reordering-based parallel method for EVC tailored to the state vector distribution (Section~\ref{sec:data_layout}).
Figure~\ref{fig:parallel_evc} provides examples of qubit reordering-based parallel EVC.
This method computes the expectation value through qubit reordering communication, ensuring Pauli strings have narrow access, akin to QSU.
Consequently, the overall qubit reordering cost relies on the application order of operations and qubit mapping.

In our parallel EVC, $p$ PEs locally compute partial expectation values for each Pauli string and these $p$ local values are reduced into the single value with an all-to-all communication.
In cases where a Pauli string $P_i$ has narrow access, our approach capitalizes on the characteristic that a Pauli string with narrow access can be transformed into the tensor product of $I^{\otimes n_G}$ and the partial Pauli string $P_i^\prime$:
\begin{align}
    \bra{\psi}P_i\ket{\psi} &=\begin{bmatrix} \bra{\psi_0} \ldots \bra{\psi_{p-1}} \end{bmatrix} I^{\otimes n_G}\otimes P_i^\prime \begin{bmatrix} \ket{\psi_0} \\ \vdots \\ \ket{\psi_{p-1}} \end{bmatrix} \notag \\
    &=\sum_{j=0}^{p-1}\bra{\psi_j}P_i^\prime\ket{\psi_j},\label{eq:parallel_evc}
\end{align}
where $0 \leq i<m$.
As a result, PE $j$ is allowed to locally compute the partial expectation value $\bra{\psi_j}P_i^\prime\ket{\psi_j}$ of $\bra{\psi}P_i\ket{\psi}$ because memory access is closed in each PE.
On the other hand, in cases where a Pauli string $P_i$ has wide access, the method performs qubit reordering such that the Pauli string has narrow access.
Thus, PE $j$ computes the local summation of Pauli strings $E_j$ while performing multiple qubit reorderings for eliminating wide access:
\begin{align}
    E_j&=\sum_{i=0}^{m-1} a_i \bra{\psi_j}P_i\ket{\psi_j}, \label{eq:local_evc}
\end{align}
where $0\leq i<m$.
At the end of the process, the proposed method produces the expectation value of the Hamiltonian by reducing the local values $E_0, E_1, \ldots, E_{p-1}$ into the single value:
\begin{align}
    \bra{\psi}H\ket{\psi}=\sum_{j=0}^{p-1} E_j, \label{eq:aggregation}
\end{align}
which requires an all-to-all communication for reduction.

The proposed method enhances the locality of qubit references to reduce communication.
The method is allowed to compute expectation values without qubit reorderings if consecutive Pauli strings have the same target qubits within the same qubit mapping.
This feature is an advantage over an existing method~\cite{mpi-qulacs} that computes expectation values by state-vector inner product.

\subsubsection{Tiling Pauli strings}

\begin{algorithm}[t]
\caption{Tiling Pauli strings for EVC.}
\label{alg:tiling_hamiltonian}
\KwIn{set $Q=\{0,1,\ldots,n-1\}$ of qubits, ordered set $C=\{P_0,P_1,\ldots,P_{m-1}\}$ of quantum gates, target map $T:C\rightarrow 2^Q$, and possible set $Q_\mathrm{locals}$ of $n_L$ local qubits.}
\KwOut{$t$ tiles $G_0,G_1,\ldots,G_{t-1}$ of Pauli strings.}
\SetKwProg{function}{Function \algoname{TilingPauliStrings}}{}{end}

\function{$(Q,C,T,Q_\mathrm{locals})$}{
        $C_\mathrm{remain} \gets C$\;
        $t\gets 0$ \Comment*[r]{Tile index}
        \While{$C_\mathrm{remain} \neq \emptyset$}{
            $n_{\max} \gets 0$\;
            $Q_{\max} \gets \emptyset$\;
            \For(\Comment*[f]{For every qubit mapping}){$Q_L\in Q_\mathrm{locals}$}{
                $G \gets \{P\in C_\mathrm{remain} \mid T(P) \subseteq Q_L\}$ \Comment*[r]{Generate a tile}
                \If(\Comment*[f]{Update the largest tile}){$n_{\max} < |G|$}{
                    $n_{\max} \gets |G|$\;
                    $Q_{\max} \gets Q_L$\;
                }
            }
            $G_t \gets \{P\in C_\mathrm{remain} \mid T(P) \subseteq Q_{\max}\}$ \Comment*[r]{Select the largest tile}
            $C_\mathrm{remain} \gets C_\mathrm{remain} \setminus G_t$ \Comment*[r]{Eliminate the Pauli strings in the selected tile}
            $t\gets t+1$
        }
        \Return $G_0,G_1,\ldots,G_{t-1}$\;
}
\end{algorithm}

The proposed method classifies $m$ Pauli strings into $t$ tiles to minimize the occurrence of qubit reorderings.
This minimization problem aligns with maximizing the tile size, i.e., the number of Pauli strings within each tile, because EVC can be processed in an arbitrary order.
Thus, greedily constructing tiles in order from largest to smallest is the optimal solution for this objective problem.
Note here that maximizing tile size translates to employing lazy qubit reordering techniques.

Algorithm~\ref{alg:tiling_hamiltonian} shows the tiling method for EVC.
The proposed method selects the tiles in descending order of size from all possible combinations (line~12).
This selection iterates until all Pauli strings are eliminated from the working set $C_\mathrm{remain}$ (line~4).
The time complexity of this algorithm is given by $\mathcal{O}\left(mt\binom{n}{n_G}\right)$ because the while loop at line~4 iterates $t$ times, the for loop at line~7 iterates $\binom{n}{n_G}$ times, and tile generation at line~8 investigates $m$ Pauli strings.
Note that the number $m$ of Pauli strings is given by at most $n^4$~\cite{tilly22pr}, which is smaller than $2^n$, or the length of the state vector. Therefore, the scheduling overhead for large $n$ is negligible compared to the subsequent EVC simulation.

Once $t$ tiles are generated by Algorithm~\ref{alg:tiling_hamiltonian}, the proposed method alternates between computing the expectation value for the current tile and performing a qubit reordering for the next tile.
As a result, qubit reordering-based parallel EVC computes the expectation value of the Hamiltonian with a reduction communication and $(t-1)$ qubit reorderings.

Moreover, we accelerate the tiling method at the following two points because the scheduling time increases with parallelism.
First, we introduce a preprocessing step, which merges Pauli strings with identical target qubits to reduce the number of Pauli strings for tiling.
Given two Pauli strings $P_i$ and $P_j$, the preprocessing step merges $P_i$ to $P_j$ if $T(P_i)\subseteq T(P_j)$.
As a result, both $P_i$ and $P_j$ are included in the same tile.
Reducing the number of Pauli strings accelerates the tiling method by a constant factor.

Regarding the second point, we parallelize the tiling method with $p~(=2^{n_G})$ PEs.
Because the while loop at line~7 limits the performance of Algorithm~\ref{alg:tiling_hamiltonian}, we parallelize the loop to accelerate finding the largest tile.
In this approach, the search space of $\binom{n}{n_G}$ qubit mappings is equally distributed to PEs, allowing PEs to independently find the locally largest tile from the assigned $\binom{n}{n_G}/2^{n_G}$ qubit mappings.
Subsequently, an all-to-all communication such as MPI\_Allgather is used to select the maximum solution from $p$ local results.
Consequently, this improved method accelerates tiling process to $p$ times faster with the overhead of $(t-1)$ all-to-all communications.

\subsubsection{Diagonalization of Pauli strings}
Diagonalization of Pauli strings is useful for reducing the occurence of qubit reorderings.
This optimization is available if a Pauli string $P$, which may possess wide access, satisfies the following condition:
\begin{align}
    \exists P_G \in \{I,Z\}^{\otimes n_G} \exists P_L \in \{I,X,Y,Z\}^{\otimes n_L}\,[P=P_G\otimes P_L]. \label{eq:without_QR}
\end{align}
This condition implies that a sub string $P_G$ affects only global qubits, allowing us to compute the expectation value for $P$ without resorting to qubit reordering communication.
Because the tensor product of $I$ and $Z$ forms a diagonal matrix, PEs can independently compute the expectation value of $P$ as follows:
\begin{align}
    \bra{\psi}P\ket{\psi} &=
    \bra{\psi}P_G\otimes P_L\ket{\psi} \notag\\
    &=\bra{\psi} \begin{bmatrix}
        s_0 & & 0\\
        & \ddots & \\
        0 & & s_{p-1}
    \end{bmatrix}\otimes P_L \ket{\psi} \notag \\
    &=\sum_{j=0}^{p-1} s_j\bra{\psi_j}P_L\ket{\psi_j}, \label{eq:par_exp}
\end{align}
where $s_j\in \{1,-1\} (0 \leq j<p)$ denote the diagonal components of $P_G$.

Combining the tiling method and diagonalization reduces the frequency of qubit reorderings because the tiling method can disregard Pauli $Z$.
Specifically, this combined method can create tiles solely for Pauli $X$ and $Y$, irrespective of the qubits affected by Pauli $Z$.
Consequently, the diagonalization increases the degree of freedom for the tiling method compared to tiling for $X, Y, $ and $Z$.
Thus, the proposed method reduces the number of tiles and decreases the occurrence of qubit reorderings.

Moreover, the proposed method is effective for second quantized Hamiltonians employing the Jordan-Wigner transformation~\cite{jordan-wigner}.
Such Hamiltonians predominantly consist of Pauli $Z$ with a maximum of four $X$ or $Y$ operators.
Therefore, the proposed method constructs tiles in a manner that confines only four $X$ or $Y$ operators to local qubits, regardless of $Z$ operators.

\section{Evaluation}
We assess the proposed methods using VQE simulation from the following perspectives.
\begin{itemize}
    \item Ablation study on simulation performance obtained with our solution $(g, M)$.
    \item Comparison of simulation performance with existing simulators.
    \item Scheduling time required for computing $(g, M)$.
\end{itemize}

\subsection{Experimental setup}

We targeted a VQE for H\textsubscript{2}O molecules using the GateFabric ansatz~\cite{gatefabric} to evaluate QSU and EVC.
We set computational precision at the double precision required for VQE.
For QSU evaluation, we employed the GateFabric circuit.
Qubit reordering reduction for the GateFabric circuit presents a challenge because the circuit uniformly affects all qubits with slight circuit depth.
In contrast, reducing qubit reorderings for biased circuits that apply to only a portion of the qubits is a more straightforward task.
The deployed ansatz operates qubits in a regular manner.
One of the advantages of the GateFabric ansatz is circuit decomposition, which is useful for simplifying the circuit.
However, we directly used the operator without making any decomposition. Thus, the advantages of the GateFabric ansatz were excluded in the experiments.
Additionally, we set the number of layers to $4n$ to scale up the circuit and increase the occurrence of qubit reorderings.
To evaluate EVC, we set H\textsubscript{2}O molecules to variable $n^\prime=n/2$ orbitals and 10 electrons on the cc-pVDZ basis.
We transformed the H\textsubscript{2}O molecules into Hamiltonians using second quantization and the Jordan-Wigner transformation.

\begin{figure}
    \centering
    \includegraphics[width=\linewidth]{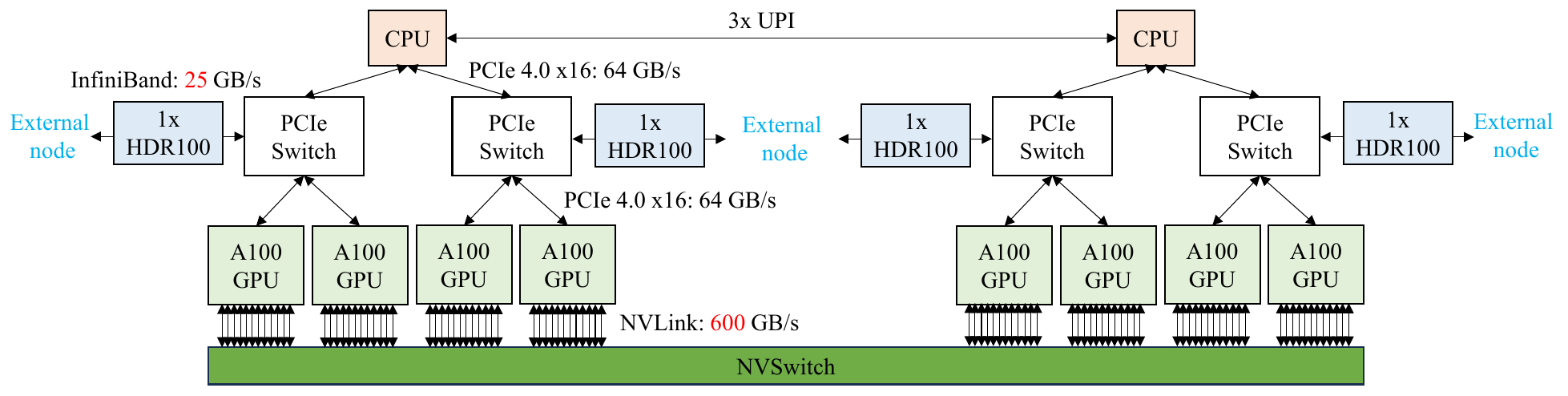}
    \caption{Semantic diagram of the GPU node in SQUID~\cite{squid}.
    Each node features two CPUs with eight NVIDIA A100 GPUs and four Infiniband cards.
    Intra-node GPU communication has a bandwidth 600 GB/s (bidirectional) facilitated by a third-generation NVSwitch.
    Inter-node communication operates at a bandwidth of 25 GB/s via InfiniBand HDR100.}
    \Description[Architectural diagram of a SQUID GPU node.]{This is an architectural diagram of a GPU node in the SQUID supercomputer. It highlights the hierarchical interconnect, showing eight GPUs connected via a high-speed NVSwitch (for intra-node communication) and InfiniBand cards for connecting to other nodes (inter-node communication).}
    \label{fig:node_block}
\end{figure}

Experiments were conducted on the supercomputer system SQUID~\cite{squid} at the D3 Center, The University of Osaka.
The system was operated with a batch system to ensure exclusive execution, which provides users a clean set of computing nodes. The batch system is the NEC Network Queuing System V, which is a variant of the well-known Portable Batch System (PBS).
SQUID is a hybrid cluster system comprising general-purpose CPU nodes, GPU-equipped nodes (GPU nodes), and vector processor nodes. GPU nodes are interconnected via InfiniBand, forming a full-bisection fat-tree topology.
Figure~\ref{fig:node_block} and Table~\ref{tab:system} show the internal network and system configuration of a single GPU node, respectively.
Consequently, the entire GPU constitutes a two-layered interconnection network.
Notably, the bandwidth of intra-node GPU-to-GPU communication is 24 times higher than that of inter-node GPU-to-GPU communication.

As for setting of parameters, $n_L$, $n_{\mathrm{intra}}$, and $n_{\mathrm{inter}}$, the basic strategy is to maximize the numbers of local and intra-node qubits as long as the memory capacity permits.
Firstly, $n_L$ (local qubits) must be maximized to avoid communication between GPUs.
Secondly, we maximize $n_{\mathrm{intra}}$ (intra-node qubits) to have less inter-node communication.
For example, we first set $n_L=31$ for 32GB of memory per GPU, $n_{\mathrm{intra}}=3$ for 8 GPUs per node, and finally, the remaining qubits are assigned to inter-node qubits: $n_{\mathrm{inter}}=n-n_L-n_{\mathrm{intra}}$.

\begin{table}[t]
    \caption{System configuration of the SQUID GPU node.}
    \label{tab:system}
    \centering
    \begin{tabular}{ll}
        \hline
        Item        & Specification                                                   \\ \hline
        CPU          & Intel Xeon Platinum 8368 (IceLake, 2.4 GHz, 38 cores) $\times$ 2 sockets\\
        RAM          & 512 GB                                                          \\
        GPU          & NVIDIA HGX A100 (40 GB VRAM) $\times$ 8 cards\\
        Interconnect & InfiniBand HDR100 $\times$ 4 cards\\
        OS           & Rocky 8.6                                                       \\ \hline
    \end{tabular}
\end{table}

We implemented the proposed methods using the cuStateVec v1.5.0 library included in cuQuantum v23.10~\cite{cuquantum_ver}.
For QSU, we utilized the \texttt{custatevecApplyMatrix} API to apply general quantum gates.
For EVC, we employed the \texttt{custatevecComputeExpectationsOnPauliBasis} API to compute the expectation value of Pauli strings.
Additionally, p-SVQCS conducted qubit reordering communication using the \texttt{custatevecSVSwapWorkerExecute} API.
GPU Direct P2P communication was enabled using the \texttt{custatevecSVSwapWorkerSetSubSVsP2P} API to enhance intra-node communication.
As for time measurement, we used the \texttt{cudaEventElapsedTime} function to obtain the median of three executions. This function has a resolution of around 0.5 $\mu$s \cite{cuda}.

The proposed method was developed in C++20 and executed from Python 3.10.13 through PyBind 2.10.3~\cite{pybind}.
To transform molecular data into the Hamiltonian format, we utilized the chemical library OpenFermion 1.5.1~\cite{openfermion}, developed with Python.
Each GPU was assigned an MPI process using OpenMPI 4.0.5 to realize inter-node parallel execution.
Note that intra-node communication was efficiently processed by shared-memory transport provided by OpenMPI.
The program was compiled using GCC 11.3.0 wrapped by \texttt{mpicxx} with the option \texttt{-Wall -O3}.
Furthermore, every process ran on a Singularity 3.7 container converted from the NVIDIA HPC SDK 22.11~\cite{nvhpc}.

\subsection{Ablation study on simulation performance}
By varying the number $n$ of qubits from 20 to 36, we evaluated the number of qubit reordering occurrences and the computation/communication time through QSU simulations on 64 GPUs ($p=64$).
Furthermore, we examined the strong scaling of the QSU simulation by varying the number $p$ of GPUs while fixing the problem size $n$. We used the maximum size ($n=30$), which can be processed on a single GPU.
Because the number $n$ of qubits is not sufficient for some GPU setups, we first set $n_{\mathrm{intra}}$ and then $n_{\mathrm{inter}}$ according to the deployed setup. For example, we set $n_{\mathrm{intra}}=2$ for 4 GPUs per node, $n_{\mathrm{inter}}=2$ for 4 nodes, and finally $n_L=26$. We also conducted weak scaling analysis, where the problem size $n$ scales with the number $p$ of GPUs.
The initialization time was not included in the measured execution time.

\subsubsection{Quantum state update}

\begin{table}[t]
    \caption{Comparable methods deployed for ablation study of QSU simulation.}
    \label{tab:compare_qsu}
    \centering
    \begin{tabular}{lccc}
        \hline
        \multirow{2}{*}{Component} & \multicolumn{3}{c}{Method}\\ \cline{2-4}
                   & Hierachical & Flat & Ad hoc\\ \hline
        Simulation & GPU-based & GPU-based & GPU-based\\
        Scheduling & out-of-order & out-of-order & inorder\\
        Interconnection & hierarchical & flat & flat\\ \hline
    \end{tabular}
\end{table}

\begin{figure}[t]
    \begin{minipage}[b]{0.48\linewidth}
        \centering
        \includegraphics[width=\linewidth]{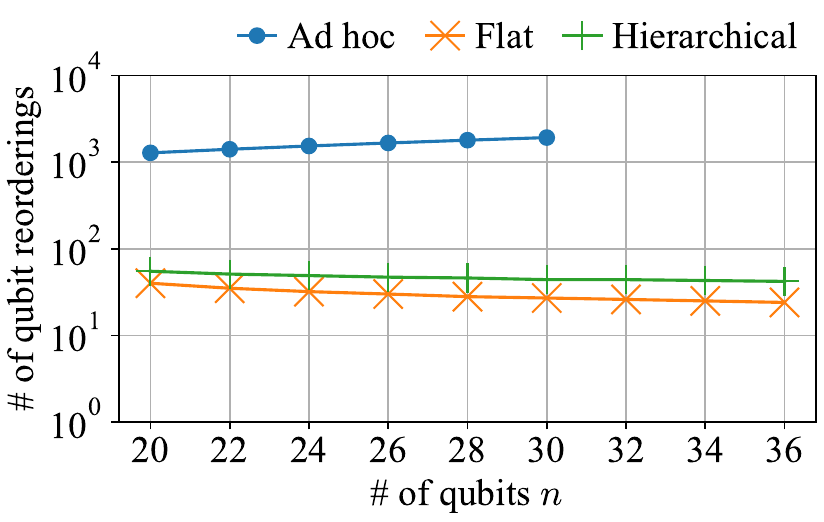}
        \subcaption{}
        \label{fig:qr_per_qubits_update}
    \end{minipage}
    \begin{minipage}[b]{0.48\linewidth}
        \centering
        \includegraphics[width=\linewidth]{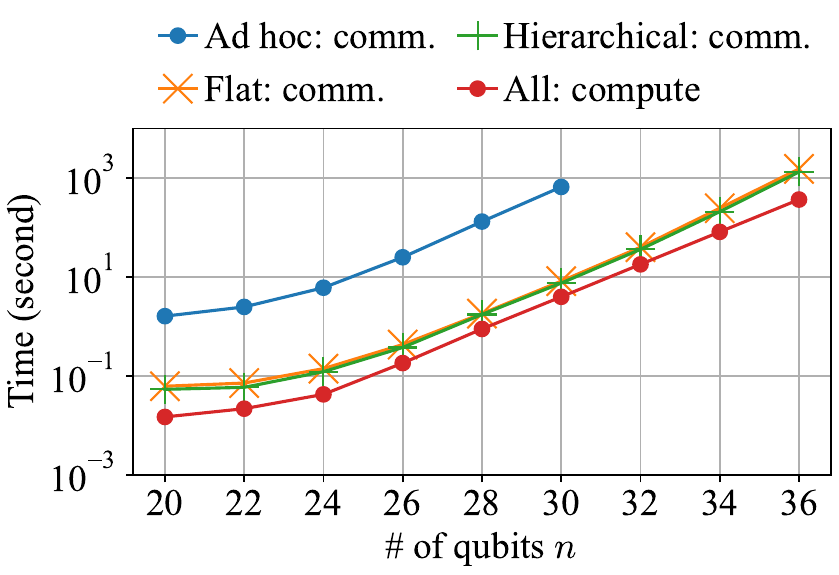}
        \subcaption{}
        \label{fig:time_per_qubits_update}
    \end{minipage}
    \begin{minipage}[b]{0.48\linewidth}
        \centering
        \includegraphics[width=\linewidth]{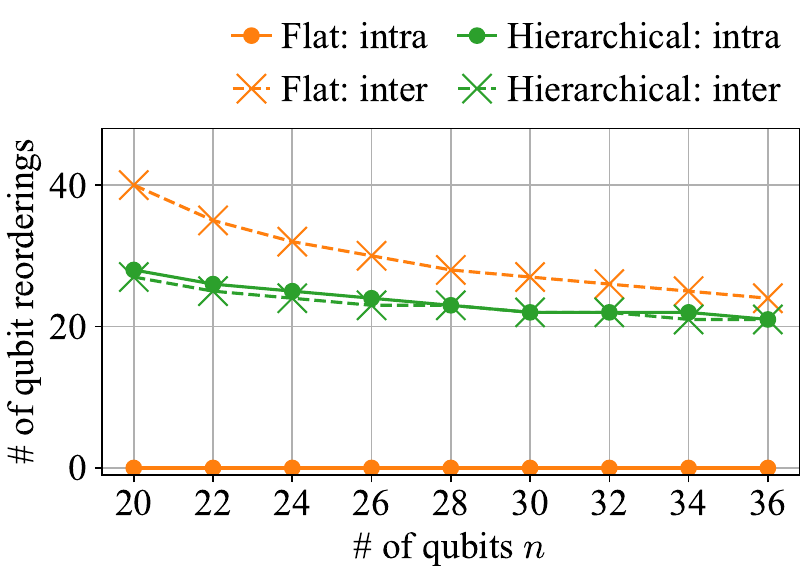}
        \subcaption{}
        \label{fig:qr_class_per_qubits_update}
    \end{minipage}
    \begin{minipage}[b]{0.48\linewidth}
        \centering
        \includegraphics[width=\linewidth]{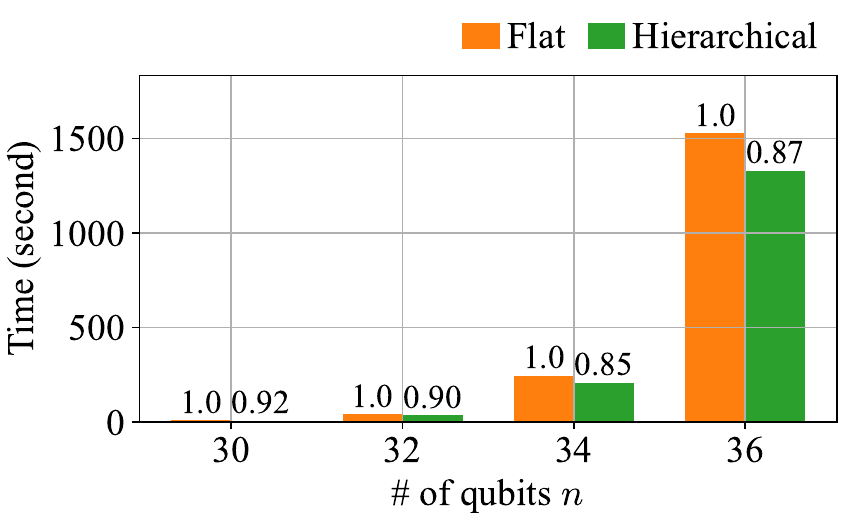}
        \subcaption{}
        \label{fig:time_per_qubits_update_comm}
    \end{minipage}
    \caption{Simulation performance of QSU utilizing the proposed methods, flat and hierarchical, on 32 GPUs ($n_{\mathrm{intra}}=3, n_{\mathrm{inter}}=2$). (a) Comparison among the ad hoc, flat, and hierarchical methods in terms of the number of qubit reorderings. (b) Breakdown of execution time. The labels ``comm.'' and ``compute'' denote communication and computation time, respectively.
    We show a single line for computation time because all methods exhibit similar computation time.
    (c) Breakdown of the number of qubit reorderings. The labels ``intra'' and ``inter'' represent the occurrences of intra-node qubit reordering and those of inter-node qubit reordering, respectively.
    (d) Comparison of communication time. The number on each bar indicates the ratio to flat; the flat value is always set as $1.0$.
    }
    \Description[QSU performance comparison of proposed methods on 32 GPUs.]{This figure contains four charts: (a) number of reorderings, (b) execution time breakdown, (c) breakdown of reordering types, and (d) communication time. Together, they demonstrate that the proposed methods (Flat, Hierarchical) significantly outperform an ad hoc method by reducing reorderings and communication time in QSU simulations.}
    \label{fig:update_performance}
\end{figure}

Table \ref{tab:compare_qsu} summarizes all of the comparable methods used for ablation study of QSU.
We compared the proposed methods with \emph{an ad hoc method}, which is a GPU-based simulator that deploys the qubit reordering mechanism of mpiQulacs v0.6.0~\cite{mpi-qulacs}. We implemented the ad hoc method by ourselves because the original mpiQulacs is a CPU-based simulator.
The ad hoc method schedules operations in a breadth-first order on the given circuit, where a gate with depth $d$ has to be executed after all gates with depth $d-1$.
If the next operation has a wide access, the ad hoc method selects a new qubit mapping involving qubit reordering communication to ensure that the next operation has a narrow access pattern without considering future operations.
Figure~\ref{fig:update_performance} summarizes the experimental runtime performance in the QSU simulation.
Here, the terms \emph{flat} and \emph{hierarchical} refer to the proposed methods described in Sections~\ref{sec:flat_tiling} and \ref{sec:hiera_tiling}, respectively.
It is worth noting that the ad hoc method was executable only with at most 30 qubits owing to time limitations.

\begin{figure}[t]
    \begin{minipage}[b]{0.33\linewidth}
        \centering
        \includegraphics[width=\linewidth]{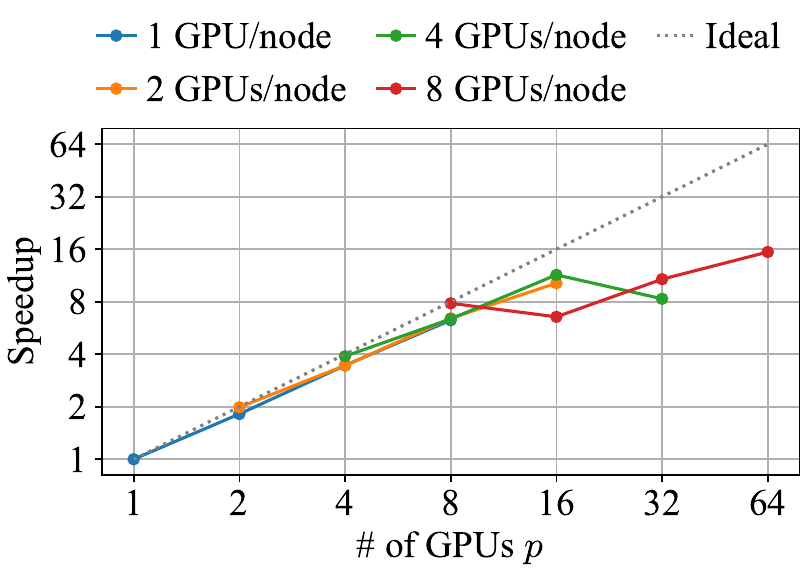}
        \subcaption{}
        \label{fig:time_scaling_update}
    \end{minipage}%
    \begin{minipage}[b]{0.33\linewidth}
        \centering
        \includegraphics[width=\linewidth]{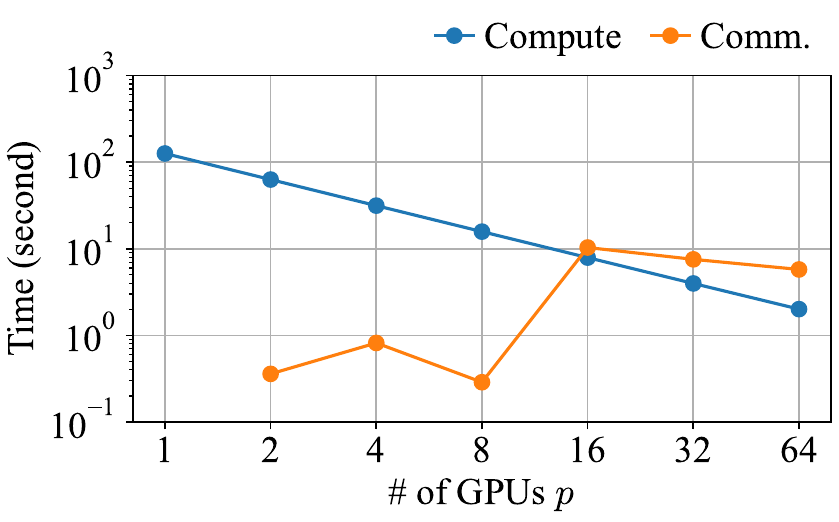}
        \subcaption{}
        \label{fig:time_scaling_update_breakdown}
    \end{minipage}%
    \begin{minipage}[b]{0.33\linewidth}
        \centering
        \includegraphics[width=\linewidth]{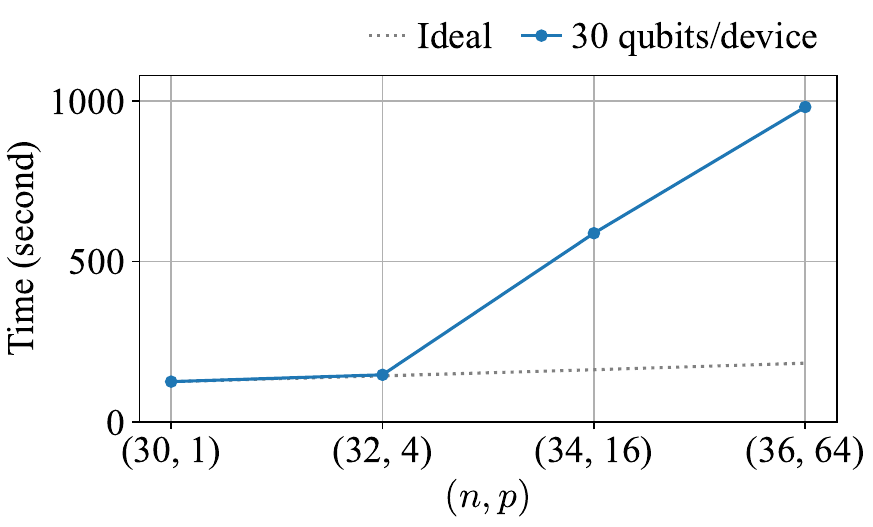}
        \subcaption{}
        \label{fig:time_weakscaling_update}
    \end{minipage}
    \caption{Analysis of the proposed hierarchical method for QSU. (a) Strong scaling for $n=30$ with different numbers of GPUs, (b) breakdown of execution time for $n=30$ with different number of GPUs, and (c) weak scaling with different numbers of qubits and GPUs. For (b) and (c), we used $n_{\mathrm{intra}}=3$ if $p \geq 8$ and $n_{\mathrm{inter}}=1$ otherwise. The dashed line in (c) represents the minimum time estimated by removing the communication time.}
    \Description[Analysis of hierarchical QSU method performance.]{This figure contains three charts analyzing QSU performance: (a) strong scaling, (b) a time breakdown of the strong scaling, and (c) weak scaling. The charts show that performance scales well within a single node but that inter-node communication becomes a bottleneck when scaling across multiple nodes.}
    \label{fig:strong_update}
\end{figure}

As shown in Fig.~\ref{fig:qr_per_qubits_update}, the proposed methods reduced qubit reordering occurrences as the number of qubits $n$ increases.
Conversely, the ad hoc method experienced an increase in the number of qubit reordering occurrences with a larger number of qubits ($\propto$ the number of layers $4n$), attributed to encountering quantum gates that require qubit reorderings for each additional depth.
In contrast, the proposed methods demonstrated a decrease in the number of qubit reordering occurrences as $n$ increases.
For instance, with 30 qubits, the flat method reduced the number of qubit reordering occurrences to 1/71 compared to the ad hoc method; the number reduced from 1920 to 27.
This reduction can be attributed to two primary factors.
First, the proposed method reorders quantum gates across the depth direction to perform qubit reorderings lazily.
Because the qubit mapping of successive gates within a tile remains consistent, the proposed method effectively suppresses the occurrence of qubit reorderings.
Second, the ratio of local qubits, $n_L/n$, increases with $n$ in the case of a fixed number of GPUs, because the number $n_G$ of global qubits depends on the number $p$ of GPUs.
As a result, the number of gates in each tile increases with $n_L/n$ because the proposed method constructs tiles that maximize the number of gates for a given $n_L$ qubits.
Therefore, the proposed method reduces the number of tiles, which leads to less occurrence of qubit reorderings particularly in scenarios involving a large number of local qubits.

Figure~\ref{fig:time_per_qubits_update} indicates that the proposed methods effectively reduced communication time without increasing the computation time regardless of $n$.
For instance, with 30 qubits, the flat method reduced communication time to 1/80 compared to the ad hoc method while decreasing the occurrence of qubit reorderings; the communication time reduced from 660.0 s to 8.2 s.
Consequently, the flat method reduced the execution time from 664.0 s to 12.2 s, accelerating QSU by a factor of 54$\times$.

Surprisingly, the hierarchical method reduced communication time further compared to the flat method despite experiencing an increase in the total number of qubit reorderings.
Detailed breakdowns of qubit reordering occurrences and communication time are shown in Figs.~\ref{fig:qr_class_per_qubits_update} and \ref{fig:time_per_qubits_update_comm}, respectively.
Intra-node qubit reordering occurrences are absent in the flat method because it tends to select the next local qubits from inter-node qubits affected by the shallow gates.
Inter-node qubit reordering occurs when at least one inter-node qubit is exchanged.
Therefore, the flat method rarely induces intra-node qubit reordering by chance, except in extreme cases where $(n_L\leq n_\mathrm{intra})$.
Conversely, the hierarchical method reduces inter-node qubit reordering occurrences in favor of prioritizing intra-node qubit reorderings.
Consequently, the hierarchical method results in more qubit reorderings than the flat method.
However, the hierarchical method still manages to decrease communication time by up to 15\% compared to the flat method primarily because the overhead of inter-node communication outweighs that of intra-node communication; the communication time reduced 245.1 s to 208.8 s.
The hierarchical method is more effective for large quantum circuits with large inter-node communication overhead.

Figure~\ref{fig:time_scaling_update} shows the strong scaling of the hierarchical method for $n=30$.
We varied the number of nodes from 1 to 8 while fixing the number of GPUs per node. The measured speedup for $p$ GPUs is given by $T_p/T_1$, where $T_p$ denote the total execution time on $p$ GPUs.
The proposed method achieved nearly linear speedup in multi-GPU execution within a single node.
The hierarchical method exhibited an almost ideal speedup on a single node, up to 8 GPUs, as depicted in Fig.~\ref{fig:time_scaling_update}.
Moreover, parallel efficiency reached levels between 0.975 and 0.989 for 8 GPU executions.
Furthermore, the hierarchical method attained maximum acceleration on multi-node setups with 64 GPUs.
However, the speedup of 16-GPU execution is inferior to that of the 8-GPU execution when using 8 GPUs per node.
This degradation was due to the inter-node communication, which never occurs on 8-GPU execution.
A similar degradation can be found between 16 GPUs and 32 GPUs with 4 GPUs per node, where the number of nodes increases from 4 to 8.
However, such a degradation cannot be found between 32 GPUs and 64 GPUs with 8 GPUs per node.
The reason for this irregular behavior can be attributed to communication, because computation time linearly reduces with the increase of $p$.
We implemented the communication with the \texttt{custatevecSVSwapWorkerExecute} function, which internally calls the \texttt{MPI\_Isend} and \texttt{MPI\_Irecv} functions to exchange all-to-all messages between GPUs.
These functions varied the communication time depending on the communication pattern and amount.
With respect to the setup of GPU resources, a promising strategy for achieving high performance is to maximize the number of GPU per node.
However, we found an exception in Fig.~\ref{fig:time_scaling_update}, where the highest performance for $p=16$ was obtained when using 4 GPUs per node rather than 8 GPUs per node.
The number of qubit reorderings was exactly same between these two setups.
Therefore, we think that this irregular behavior was caused by the communication pattern, which determines the communication time.

Figure~\ref{fig:time_scaling_update_breakdown} illustrates that the increase in communication time decreased the speedup.
Ideally, the computation time decreases as $p$ increases.
However, the communication time significantly increased from 8-GPU to 16-GPU execution owing to message passing among 16 GPUs transitioning from the intra-node communication to slower inter-node communication.
When message passing remains confined within the same interconnection layer, communication time is assumed to be uniform regardless of $p$.
This is realized by the underlying cuQuantum library, which is based on the distributed multiple pair communication~\cite{cuquantum}.

Figure~\ref{fig:time_weakscaling_update} shows the weak scaling of the hierarchical method, where $n$ was increased with $p$.
We increased $p$ by a factor of 4 each time because our simulator accepted only even values of $n$.
We set $n_{\mathrm{inter}}=1$ if $p\leq4$
and $n_{\mathrm{intra}}=3$ otherwise.
The hierarchical method rapidly increased the execution time for $p=16$ and $64$.
The dashed line, which represents the minimum time estimated by removing the communication time, indicates that the increase was due to the inter-node communication.
A single reordering of qubits involves an all-to-all communication across $p$ GPUs, with the communication size depending on the state vector length ($2^n$).
Therefore, the communication overhead becomes substantial when $p>4$, and moreover,
the number of qubit reorderings increased with $n$, generating 19 reorderings for $(n,p)=(32,4)$ and 46 reorderings for $(n,p)=(36,64)$.

\subsubsection{Expectation value computation}

\begin{figure}[t]
    \begin{minipage}[b]{0.33\linewidth}
        \centering
        \includegraphics[width=\linewidth]{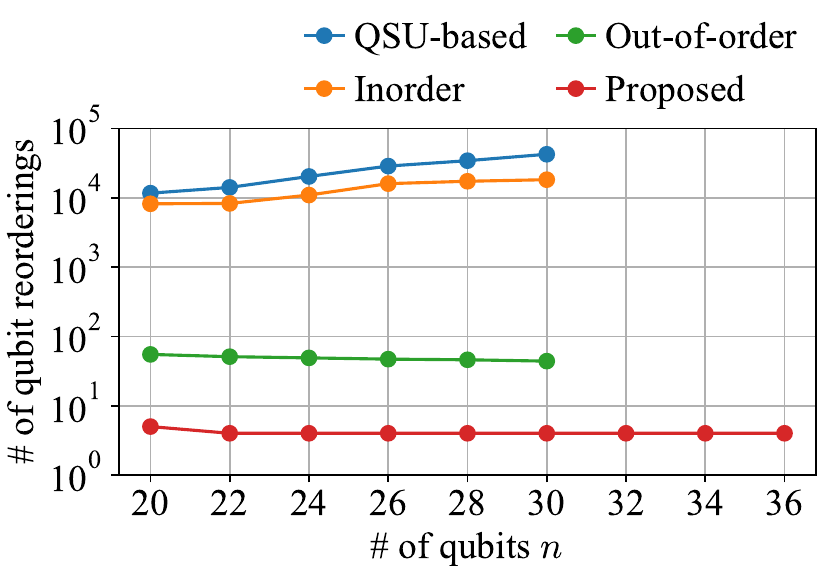}
        \subcaption{}
        \label{fig:qr_per_qubits_exp}
    \end{minipage}%
    \begin{minipage}[b]{0.33\linewidth}
        \centering
        \includegraphics[width=\linewidth]{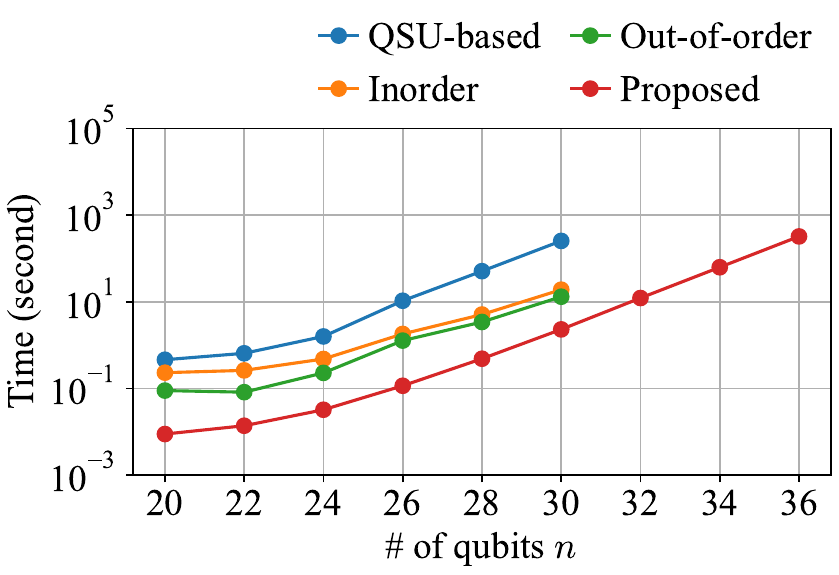}
        \subcaption{}
        \label{fig:comp_time_per_qubits_exp}
    \end{minipage}%
    \begin{minipage}[b]{0.33\linewidth}
        \centering
        \includegraphics[width=\linewidth]{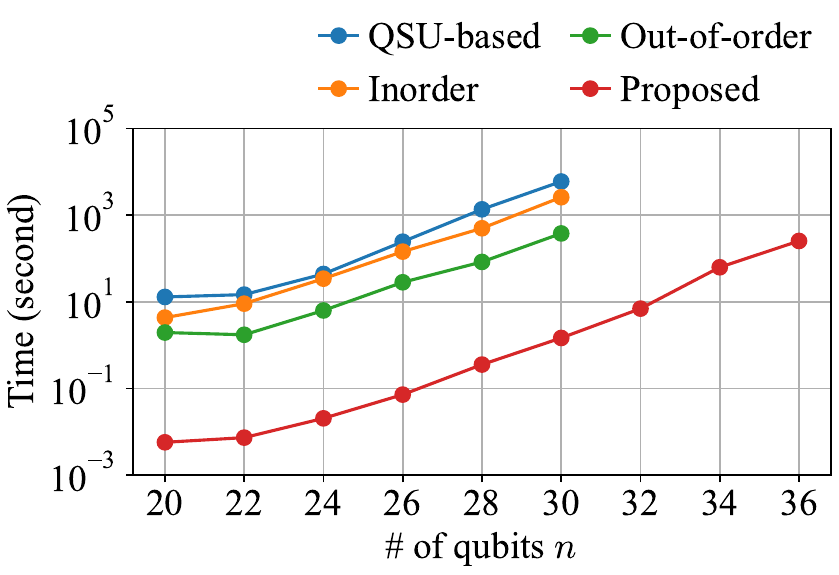}
        \subcaption{}
        \label{fig:comm_time_per_qubits_exp}
    \end{minipage}
    \caption{Simulation performance of EVC utilizing the proposed methods on 32 GPUs ($n_{\mathrm{intra}}=3, n_{\mathrm{inter}}=2$). (a) Comparison of the number of qubit reorderings, (b) that of computation time, and (c) that of communication time.}
    \Description[EVC performance comparison of proposed methods on 32 GPUs.]{This figure presents three charts on EVC performance: (a) number of reorderings, (b) computation time, and (c) communication time. They show that the proposed method is orders of magnitude faster than a QSU-based approach, dramatically reducing all three metrics.}
    \label{fig:exp_performance}
\end{figure}

Table~\ref{tab:compare_evc} summarizes the comparable methods used for ablation study of EVC.
We compared the proposed method (described in Section~\ref{sec:proposed_evc}) with the \emph{QSU-based} method, which computes $\bra{\psi}H\ket{\psi}$ using QSU with the ad hoc method.
First, the QSU-based method executes QSU to obtain $\ket{\psi^\prime} = H\ket{\psi}$ regarding $P_i$ as a quantum gate; the application order of Pauli strings is the ascending order of $i$ given by the OpenFermion transformation.
This index-ordered computation method is utilized in mpiQulacs~\cite{mpi-qulacs}.
Second, the QSU-based method employs the cuBLAS API~\cite{cublas} to compute the inner product $\braket{\psi|\psi^\prime}$.
Figure~\ref{fig:exp_performance} summarizes the experimental runtime performance in EVC.
Here, the terms \emph{inorder} and \emph{out-of-order} refer to subsets of the proposed method, as shown in Table~\ref{tab:compare_evc}.
The QSU-based method was executable with 30 or fewer qubits owing to time limitations similar to the evaluations of QSU.

\begin{table}[t]
    \caption{Comparable methods deployed for ablation study of EVC simulation. All methods assume flat interconnection.}
    \label{tab:compare_evc}
    \centering
    \resizebox{\linewidth}{!}{%
    \begin{tabular}{lcccc}
        \hline
        \multirow{2}{*}{Component} & \multicolumn{3}{c}{Method}\\ \cline{2-5}
                & Proposed & Out-of-order & Inorder & QSU-based\\ \hline
        (1) EVC & qubit reordering-based &  qubit reordering-based &  qubit reordering-based & QSU-based\\
        (2) Tiling & yes & yes & no & N/A\\
        (3) Diagonalization & yes & no & no & N/A\\ \hline
    \end{tabular}%
    }
\end{table}

As illustrated in Fig.~\ref{fig:qr_per_qubits_exp}, the proposed method reduced the number of qubit reordering occurrences from 42,512 to 4 (1/10,628) for $n=30$ by rearranging the order of Pauli strings compared to the QSU-based method.
This result underscores the significance of minimizing qubit reordering occurrences.
The proposed method required only four qubit reorderings to compute the expectation value of a large Hamiltonian comprising $\mathcal{O}(10^4)$ Pauli strings for 36 qubits.
The large gap between the inorder and out-of-order methods in Fig.~\ref{fig:qr_per_qubits_exp} indicates that, among the three components in Table~\ref{tab:compare_evc}, the out-of-order scheduling has the highest impact on the reduction of qubit reorderings.

The proposed method reduced computation and communication time compared to the QSU-based method (Figs.~\ref{fig:comp_time_per_qubits_exp} and \ref{fig:comm_time_per_qubits_exp}).
With the decrease in qubit reordering occurrences, the proposed method reduced communication time from 5,979 s to 1.46 s, accelerating the total communication by 4,094$\times$ compared to the QSU-based method.
Furthermore, it also reduced computation time from 254 s to 2.3 s, accelerating total computations by 110$\times$.
This improvement stems from the optimization of the proposed method for Pauli basis computation.
While the QSU-based method treats Pauli strings as general quantum gates and utilizes the \texttt{custatevecApplyMatrix} API for computation, the proposed method computes all the expectation values on a Pauli basis with diagonalization, thereby localizing the target qubits of Pauli strings.
Consequently, the proposed method utilizes a faster API, \texttt{custatevecComputeExpectationsOnPauliBasis}, instead of \texttt{custatevecApplyMatrix}.
As a result, the proposed method reduced the entire EVC time from 6,233 s to 3.76 s, accelerating the entire EVC by 1,657$\times$ compared to the QSU-based method.

\begin{figure}[t]
    \begin{minipage}[b]{0.33\linewidth}
        \centering
        \includegraphics[width=\linewidth]{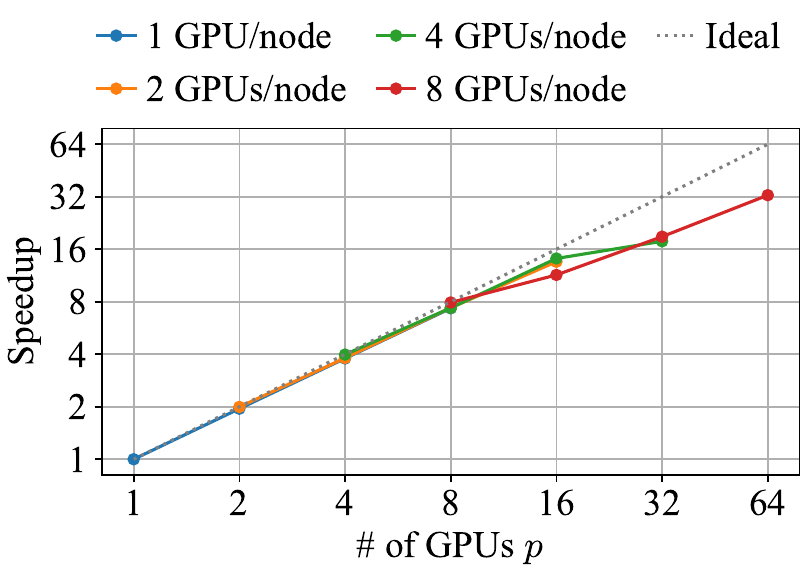}
        \subcaption{}
        \label{fig:time_scaling_exp}
    \end{minipage}%
    \begin{minipage}[b]{0.33\linewidth}
        \centering
        \includegraphics[width=\linewidth]{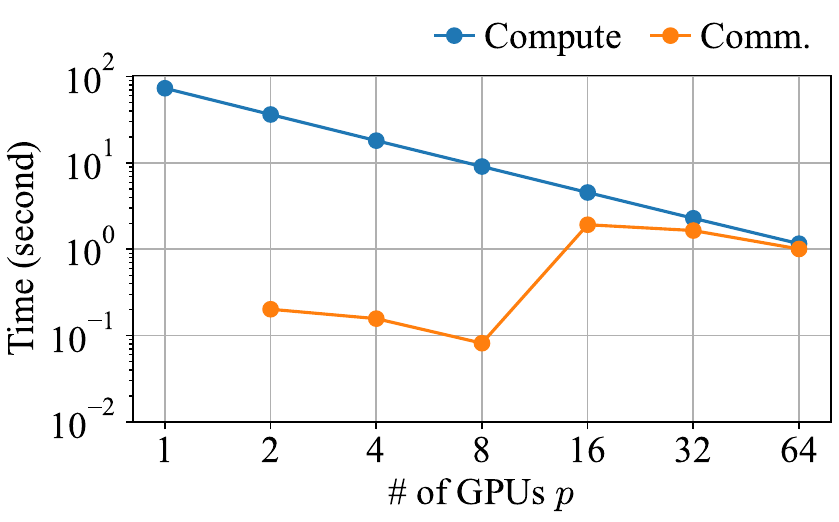}
        \subcaption{}
        \label{fig:time_scaling_exp_breakdown}
    \end{minipage}%
    \begin{minipage}[b]{0.33\linewidth}
        \centering
        \includegraphics[width=\linewidth]{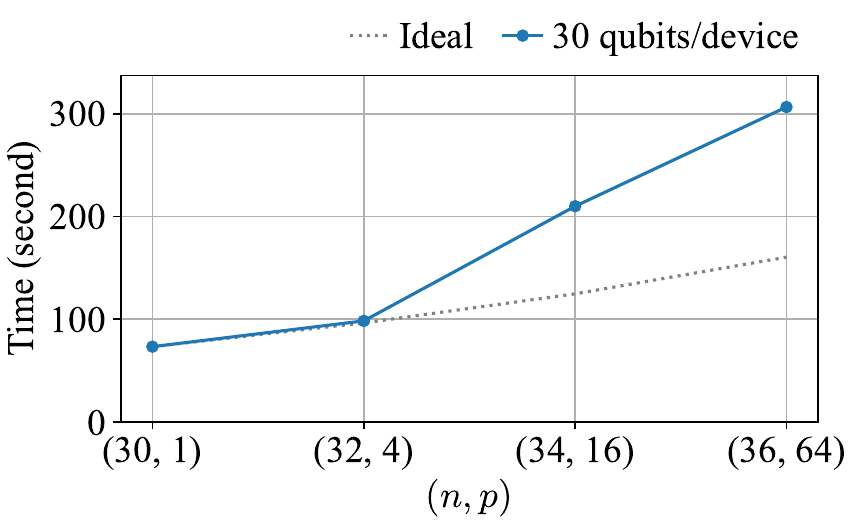}
        \subcaption{}
        \label{fig:time_weakscaling_exp}
    \end{minipage}
    \caption{Analysis of the proposed method for EVC. (a) Strong scaling for $n=30$ with different numbers of GPUs, (b) breakdown of execution time for $n=30$ with different numbers of GPUs, and (c) weak scaling with different numbers of qubits and GPUs. For (b) and (c), we used $n_{\mathrm{intra}}=3$ if $p \geq 8$ and $n_{\mathrm{inter}}=1$ otherwise. The dashed line in (c) represents the minimum time estimated by removing the communication time.}
    \Description[Analysis of proposed EVC method performance.]{This figure contains three charts analyzing EVC performance: (a) strong scaling, (b) a time breakdown of the strong scaling, and (c) weak scaling. The charts demonstrate that the proposed method for EVC has very low communication overhead and achieves nearly ideal parallel efficiency.}
    \label{fig:strong_exp}
\end{figure}

The proposed method for EVC demonstrated an ideal linear speedup in multi-GPU execution within a single node (Fig.~\ref{fig:time_scaling_exp}).
In the case of execution for 30 qubits on 8 GPUs, the proposed method achieved a speedup of 7.99 with a parallel efficiency of 0.999.
The attainment of ideal performance can be attributed to the negligibly small communication overhead in the proposed method during single node execution (Fig.~\ref{fig:time_scaling_exp_breakdown}).
Specifically, the proposed method increased the computation-to-communication ratio to as high as 99.4\% owing to the utilization of high-bandwidth NVSwitch in single-node executions.
During multi-node executions, the parallel efficiency of the proposed method decreased owing to the significant overhead of inter-node communication via InfiniBand, similar to the case of QSU.
However, the proposed method increased the computation-to-communication ratio even more when targeting EVC compared to QSU, resulting in a substantial speedup.

As we did for QSU, we also conducted the weak scaling analysis for EVC using H\textsubscript{2}O molecules (Fig.~\ref{fig:time_weakscaling_exp}).
Although the execution time increased with $p$, the increase in Fig.~\ref{fig:time_weakscaling_exp} is lower than that in Fig.~\ref{fig:time_weakscaling_update}.
This improved behavior can be attributed to diagonalization of Pauli strings, which reduced communication time, as shown in Fig.~\ref{fig:comm_time_per_qubits_exp}.
Note here that the ideal time in Fig.~\ref{fig:time_weakscaling_exp}, which excluded the communication time, increased with $n$.
This behavior can be explained by the number of Pauli strings, which increases with $n$.
The number of Pauli strings increased from 21,192 $(n=30)$ to 46,367 $(n=36)$, which increased the ideal time by 2.19$\times$.

\subsection{Comparison of simulation performance}

\begin{table}[t]
    \caption{Molecules used for comparison of simulation performance.}
    \label{tab:molecule}
    \centering
    \begin{tabular}{lcccccccc}
        \hline
        \multirow{2}{*}{Item} & \multicolumn{8}{c}{Molecule}\\ \cline{2-9}
             & N$_2$ & C$_2$H$_4$ & CO$_2$ & O$_3$ & C$_2$H$_6$ & CHOOH & Cl$_2$ & NaCl\\ \hline
        \# of qubits & 20 & 28 & 30 & 30 & 32 & 34 & 36 & 36\\
        \# of electrons & 14 & 16 & 22 & 24 & 18 & 24 & 34 & 28 \\
        \# of Pauli strings & 2,239 & 8,919 & 11,434 & 20,554 & 24,921 & 70,564 & 22,487 & 43,379\\ \hline
    \end{tabular}
\end{table}

We compared the proposed methods with QuEST v4.0.0~\cite{quest}, original mpiQulacs v0.6.0~\cite{mpi-qulacs}, and Qiskit Aer 0.17.0~\cite{qiskit-aer} by using multiple molecules: N$_2$, C$_2$H$_4$, CO$_2$, O$_3$, C$_2$H$_6$, CHOOH, Cl$_2$, and NaCl, with the STO-3G basis.
The original mpiQulacs is a CPU-based simulator and the others are GPU-based simulators.
The proposed method and QuEST were executed using the same GPU setup.
As for Qiskit Aer, its multi-threaded CPU version returned the correct results, but we could not obtain results on GPUs.
Consequently, we used the CPU version of Qiskit Aer for experiments.
Table~\ref{tab:molecule} summarizes the molecules downloaded from the Computational Chemistry Comparison and Benchmark DataBase (CCCBDB)~\cite{cccbdb}.

\begin{figure}[t]
    \begin{minipage}[b]{0.48\linewidth}
        \centering
        \includegraphics[width=\linewidth]{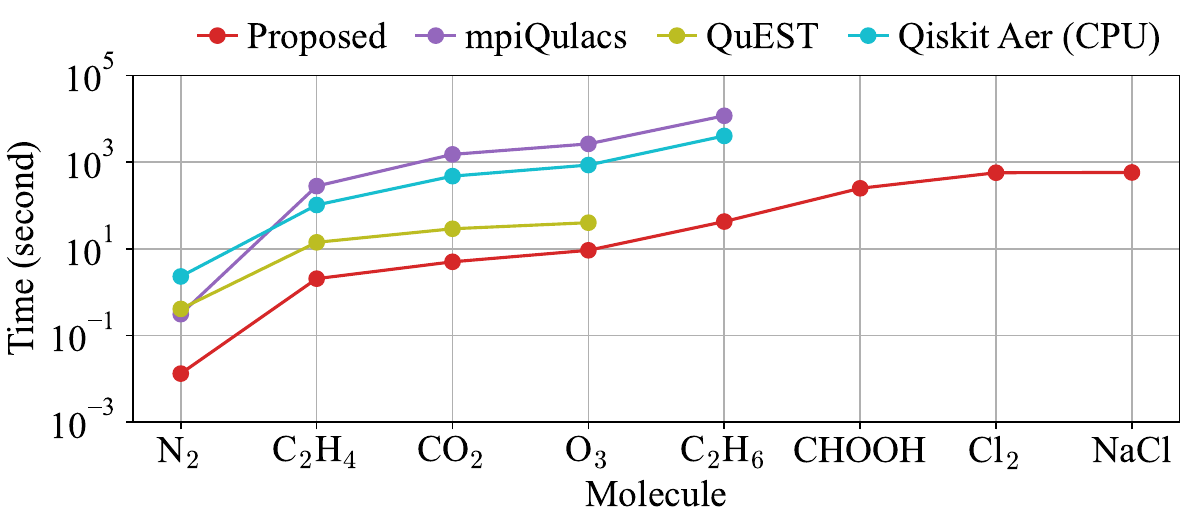}
        \subcaption{}
        \label{fig:compare_mol_time}
    \end{minipage}
    \begin{minipage}[b]{0.48\linewidth}
        \centering
        \includegraphics[width=\linewidth]{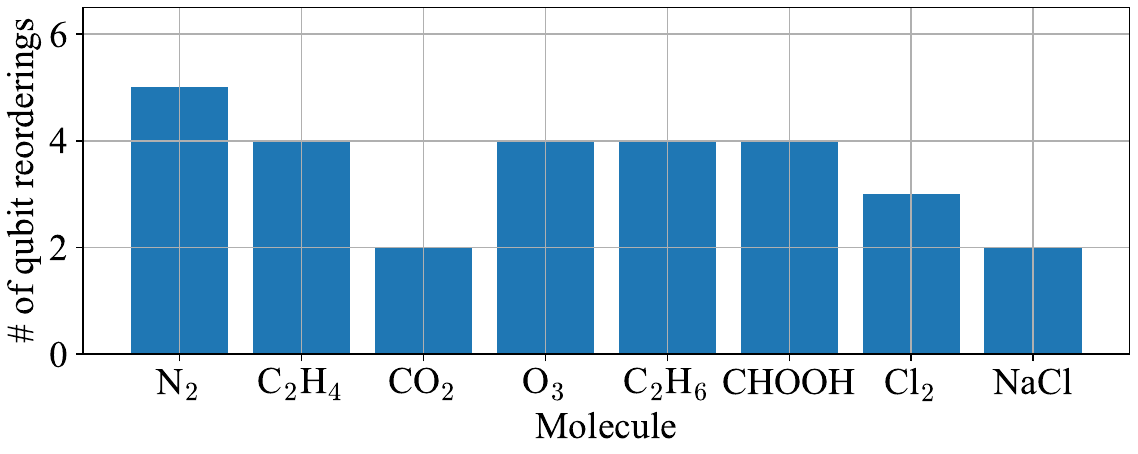}
        \subcaption{}
        \label{fig:compare_mol_qr}
    \end{minipage}
    \caption{Simulation performance with different molecules on 32 GPUs with 4 nodes ($n_{\mathrm{intra}}=3, n_{\mathrm{inter}}=2$) or only 8 CPU sockets of the same 4 nodes. (a) Measured time for comparable methods and (b) number of qubit reorderings processed by the proposed method. We failed to obtain results for C2H6, CHOOH, Cl$_2$, and NaCl due to memory exhaustion (QuEST) and long execution time of more than $10^5$ s (mpiQulacs). We show the CPU version of Qiskit Aer, because it failed to execute EVC simulation on the GPU.}
    \Description[EVC performance comparison across different molecules.]{This figure has two bar charts. (a) compares the EVC execution time of the proposed method against other simulators (mpiQulacs, QuEST, Qiskit Aer) for various molecules, showing the proposed method is fastest. (b) shows that the proposed method required five or fewer reorderings for all molecules tested.}
    \label{fig:compare_mol}
\end{figure}

Figure~\ref{fig:compare_mol_time} shows the simulation performance for different molecules on 32 GPUs and 8 CPUs with 4 nodes ($n_{\mathrm{intra}}=3, n_{\mathrm{inter}}=2$).
As compared with QuEST, the proposed method reduced the execution time for N$_2$ molecules from 0.41 s to 0.013 s, achieving a speedup of 31$\times$.
Similarly, the speedup over the original mpiQulacs ranged from 24$\times$ to 302$\times$: the lowest and highest speedups were observed for N$_2$ and CO$_2$ molecules (0.31 s over 0.013 s and 1510.2 s over 5.0 s), respectively.
Although the original mpiQulacs is a CPU-based simulator, the original paper~\cite{mpi-qulacs} reported the highest performance for large-scale simulation among GPU-based, older versions of Qiskit Aer and QuEST.
Thus, our out-of-order approach demonstrated competitive performance for large-scale simulation on tens of GPUs.

Figure~\ref{fig:compare_mol_qr} shows the number of qubit reorderings processed by the proposed method.
Similar to H$_2$O molecules, the proposed method significantly reduced the number of qubit reorderings to at most five.
This reduction allows us to deal with large-scale molecules such as Cl$_2$ and NaCl, which took around 10 minutes on 32 GPUs.
Thus, the number of qubit reorderings ranged from two to five, demonstrating the advantage of our out-of-order approach.

\begin{figure}[t]
    \centering
    \includegraphics[width=.5\linewidth]{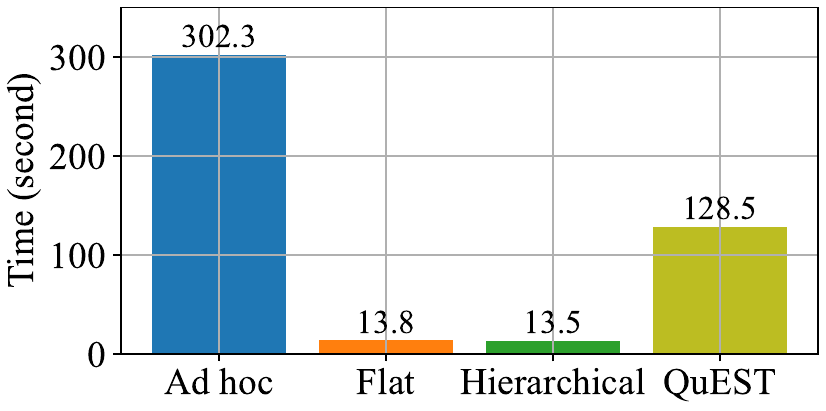}
    \caption{Simulation performance for quantum volume circuit ($n=30,d=100$) on 32 GPUs ($n_{\mathrm{intra}}=3, n_{\mathrm{inter}}=2$). mpiQulacs and Qiskit Aer failed to obtain results due to a runtime error.}
    \Description[Simulation performance for an irregular quantum volume circuit.]{This is a bar chart comparing simulation time for an irregular quantum volume circuit. It shows that the proposed methods (Flat and Hierarchical) are significantly faster than both the ad hoc method and the QuEST simulator.}
    \label{fig:random}
\end{figure}

Figure~\ref{fig:random} shows the simulation performance for a quantum volume circuit ($n=30$) that has an irregular pattern in terms of the qubits to be operated.
This circuit consisted of 100 layers, where each layer had $n/2$ 2-qubit unitary gates randomly sampled with a Haar-random distribution.
Since the circuit involved randomness, time measurements were performed 100 times to obain the median value.
Compared with the ad hoc method, our out-of-order approach reduced the execution time from 302.3 s to 13.8 s, and moreover, the hierarchical method further reduced the execution time to 13.5 s, achieving a speedup of 22$\times$.
Although this speedup is relatively lower than the speedups for QSU (54$\times$) and EVC (1,657$\times$), the out-of-order approach demonstrated a significant improvement over an inorder approach.
We also tried to run the same simulation with QuEST, mpiQulacs, and Qiskit Aer.
QuEST completed the same simulation in 128.5 s, which was faster than the ad hoc method but slower than the flat and hierarchical methods (13.8 s and 13.5 s, respectively).
On the other hand, mpiQulacs and Qiskit Aer could not complete the simulation due to a runtime error.
In summary, we think that our out-of-order approach is useful to accelerate large-scale simulation on tens of GPUs.

\subsection{Scheduling time}

\begin{figure}
    \centering
    \includegraphics[width=0.55\linewidth]{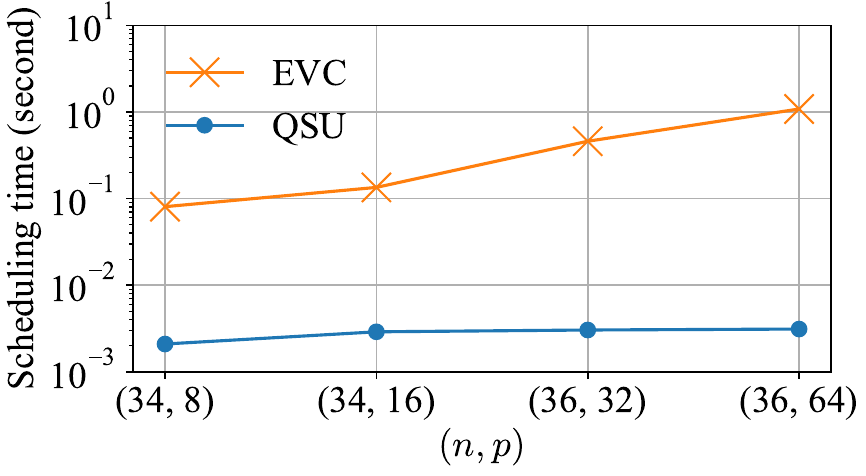}
    \caption{Scheduling time with different numbers $(n,p)$ of qubits and GPUs.}
    \Description[Scheduling time overhead for QSU and EVC.]{This chart shows the time cost of the initial scheduling step for both QSU and EVC. It demonstrates that this overhead is negligible (less than 1 second for EVC and 0.01 seconds for QSU) compared to the total simulation time.}
    \label{fig:scheduling_time}
\end{figure}

Figure~\ref{fig:scheduling_time} illustrates the scheduling time, which includes the initialization of the GateFabric ansatz, required to compute $(g, M)$ for QSU and EVC.
The scheduling time was relatively small compared to the execution time of QSU and EVC.
For QSU, the scheduling time was consistently less than $10^{-2}$ seconds across all cases.
This is attributed to the proposed scheduling algorithm for QSU, which has a time complexity of $\mathcal{O}(\overline{d}m)$ (Section~\ref{sec:flat_tiling}).
Consequently, the scheduling process is negligible when compared to the subsequent QSU simulation, which exhibits exponential time complexity with respect to $n$.
Similarly, scheduling time for EVC remains under one second across all cases.
Although the scheduling time for EVC is larger than that for QSU owing to its algorithm's time complexity of $\mathcal{O}\left(mt\binom{n}{n_G}\right)$, the time remains much smaller compared to the gain from the proposed method.
For example, the scheduling time (0.46 s) for 36 qubits on 32 GPUs accounted for only 0.08\% of the total EVC execution time (576.0 s).

Consequently, the additional scheduling process facilitated by the proposed methods has a negligible impact on the overall VQE process.
Since VQE iterates the simulation of the same circuit, the solution $(g, M)$ can be reused across all iterations.
This allows us to leverage the proposed methods without incurring significant temporal loss.
Similarly, the proposed methods are useful for various quantum applications involving iterative simulation using the same circuits, such as sampling noisy quantum circuits.

\section{Conclusion}
In this study, we have presented acceleration methods for parallel quantum circuit simulation based on state-vector methods.
The proposed methods enhance the execution order of quantum operations in two time-intensive procedures in VQE, namely QSU and EVC.
By employing time-space tiling to a set of quantum operations, the proposed methods facilitate lazy qubit reordering, thereby increasing computational granularity.
For QSU, we introduced hierarchical tiling that accounts for the interconnection of a cluster of multi-GPU computers.
In the case of EVC, we proposed three techniques: (1) parallel qubit reordering-based EVC, (2) tiling of Pauli strings, and (3) diagonalization of Pauli strings to minimize the number of tiles.

The experimental results demonstrate that the proposed methods effectively accelerated QSU and EVC by reducing the occurrence of qubit reorderings on a cluster of multi-GPU computers.
In QSU simulations involving a 30-qubit GateFabric circuit, the proposed flat tiling method reduced the number of qubit reorderings to 1/71, leading to a 54$\times$ speedup compared to the ad hoc method.
For the case of a two-layered interconnected cluster system, the hierarchical tiling method further reduced communication time by up to 15\% compared to the flat tiling method.
In EVC simulations for 30-qubit H\textsubscript{2}O molecules, the proposed method decreased the number of qubit reorderings to 1/10,628, achieving a 1,657$\times$ speedup compared to the method that computes the expectation values in the index order of the Pauli strings with performing ad hoc qubit reorderings.
Additionally, we compared the proposed methods with existing simulators and confirmed that the proposed methods achieved high performance for large-scale simulation on tens of GPUs.
These findings underscore the utility of the proposed methods for facilitating large-scale quantum circuit simulations on massively parallel machines such as supercomputers.

Future work is to enlarge a feasible problem size of VQE simulations.
Leveraging our computational method using the distributed state vector, we can expedite the search for the ground state of molecules.
Using large-scale VQE simulations, we elucidate the convergence behavior and computing resource requirements in VQE thereby paving the way for future implementations on real quantum computers.
Considering realistic noise models is also left as a future work.

\begin{acks}
This study was supported in part by the Japan Society for the Promotion of Science KAKENHI Grant Number JP22K11972, JP23K18464, JP23H03819, the MEXT Quantum Leap Flagship Program (MEXT Q-LEAP) Grant Number JPMXS0120319794, and the JST COI-NEXT program Grant Number JPMJPF2014.
This work was achieved through the use of SQUID at the D3 Center, The University of Osaka.
The authors are grateful for the anonymous reviewers for their valuable comments.
\end{acks}

\bibliographystyle{ACM-Reference-Format}
\bibliography{references}

\end{document}